\documentclass[pre,twocolumn,showpacs]{revtex4-1}

\usepackage{graphicx}
\usepackage{dcolumn}
\usepackage{lineno}
\usepackage{epsfig,amsmath,amssymb,graphics,color,calc,dcolumn,bm}

\newcommand{\ab}{{\it ab initio }}

\newcommand{\Q}[1]{Q$^{(#1)}$}

\newcommand{\boric}{{B$_2$O$_3$}}
\newcommand{\silica}{{SiO$_2$}}
\newcommand{\soda}{{Na$_2$O}}

\newcommand{\biii}{$^{[3]}$B\,}
\newcommand{\biv}{$^{[4]}$B\,}
\newcommand{\bnmr}{$^{11}$B}

\begin{document}


\title{FIRST PRINCIPLES STUDY OF A SODIUM BOROSILICATE GLASS-FORMER II: THE GLASS STATE}


\author{Laurent Pedesseau}
 \altaffiliation{present address: Universit\'e Europ\'enne de Bretagne,
INSA, FOTON, UMR 6082, 35708 Rennes}
\author{Simona Ispas}
\email[]{simona.ispas@univ-montp2.fr}
 \author{Walter Kob}
\email[]{walter.kob@univ-montp2.fr}
\affiliation{Laboratoire Charles Coulomb, UMR
5221, Universit\'e Montpellier 2 and CNRS, 34095 Montpellier, France }


\begin{abstract}

We use {\it ab initio} simulations to investigate the properties of a
sodium borosilicate glass of composition 3Na$_2$O-B$_2$O$_3$-6SiO$_2$.
We find that the broadening of the first peak in the radial distribution
functions $g_{\rm BO}(r)$ and $g_{\rm BNa}(r)$ is due to the presence of
trigonal and tetrahedral boron units as well as to non-bridging oxygen
atoms connected to BO$_3$ units. In agreement with experimental results
we find that the \biii units involve a significant number of non-bridging
oxygens whereas the vast majority of \biv have only bridging oxygens. We
determine the three dimensional distribution of the Na atoms around
the \biii and \biv units and use this information to explain why the
sodium atoms associated to the latter share more oxygen atoms with the
central boron atoms than the former units. From the distribution of the
electrons we calculate the total electronic density of states as well its
decomposition into angular momentum contributions. The vibrational density
of states shows at high frequencies a band that originates from the motion
of the boron atoms. Furthermore we show that the \biii and \biv units
give rise to well defined features in the spectrum which thus can be used
to estimate the concentration of these structural entities. The contribution
of \biii can be decomposed further into symmetric and asymmetric parts
that can also be easily identified in the spectrum. We show that certain
features in the spectrum can be used to obtain information on the type
of atom that is the second nearest neighbor of a boron in the \biv unit.
We calculate the average Born  charges on the bridging and non-bridging oxygen atoms
and show that these depend linearly on the angle between the two bonds
and the distance from the connected cation, respectively.  Finally we
have calculated the frequency dependence of the dielectric function
as well as the absorption spectra. The latter is in good quantitative
agreement with the experimental data.

\end{abstract}


\maketitle

\section{\label{sec:intro}Introduction}

Sodium borosilicate (NBS) glasses are an important class of materials
since they are resistant to thermal shocks, chemically inert, mechanically
strong, and have good insulating properties. As a consequence one
finds them in a multitude of applications such as glassware in the lab,
utensils in the kitchen, or in glass-wool for insulation. Many of these
properties are related to the way the cations Si and B build up the
network of the glass. However, since boron can form three and fourfold
coordinated structural units (i.e. triangles, \biii, or tetrahedra,
$^{[4]}$B) which have rather different mechanical and electronic
behavior, our knowledge on the global structure of the network is far
from complete~\cite{Varshneya_book}.

Using solid-state nuclear magnetic resonance spectroscopy (NMR)
of \bnmr, Yun, Bray, Dell and co-workers have investigated
intensively the structural properties of sodium borosilicate
glasses~\cite{Yun01,Yun02,Dell19831}. Based on these experiments they have
proposed a structural model (called hereafter YBD) which describes the
evolution of the structure if cations (Na) atoms are added to the system
and the mechanism of creation of non-bridging oxygens in terms of the two
parameters $K=[$\silica $]/[$\boric $]$ and $ R=[$\soda$]/[$\boric$]$,
where $[.]$ indicate mol\%. The YBD model assumes that the borosilicate
glasses contain several larger structural units like  diborate,
pyroborate, boroxol rings, reedmergnerite and danburite, which are
also referred to as supra-structural units~\cite{Varshneya_book}. These
units are in turn composed of more basic units, such as four-coordinated
silicon, three- and four-coordinated borons, etc. Within the model the
\soda-\boric-\silica~ternary diagram is divided into four compositional
regions and for each domain the YBD model predicts the fraction of
\biii and \biv~for a given boron concentration as well as the fraction
of bridging oxygens. In particular it is postulated that BO$_3$
units are transformed to BO$_4$ if $R\le 0.5$, with sodium acting as
charge-compensator, while for $0.5< R\le R^*$ open reedmergnerite-like
structures are formed on intermediate length scales. (Here the threshold
composition is given by $R^*=0.5+K/16$ and corresponds to compositions
for which the additional cations start to form non-bridging oxygen
connected to silica tetrahedra.) This latter structures consist of a
BO$_4$ tetrahedron surrounded by four SiO$_4$ tetrahedra.

Subsequently Raman spectroscopy has
been used to gain insight into the structure of NBS
glasses~\cite{Parkinson20074076,Parkinson2007415114,Parkinson20081936,Manara20092528,Manara2009777}.
For this, researchers have assigned certain bands of the spectrum to medium
range order structures but so far no consensus has emerged on how this
should be done. Based on certain features of Raman spectra measured for
sodium borosilicate glasses at different temperatures, Manara  {\it et al.}
\cite{Manara20092528,Manara2009777} have recently proposed a revision of
the YDB model. The peaks around 580 cm$^{-1}$ and 630 cm$^{-1}$, which
are two of the main features in the 550-850 cm$^{-1}$ frequency range,
have been assigned to breathing modes of borosilicate reedmergnerite-like
and danburite-like rings, respectively. (Reedmergnerite-like ring
includes one BO$_4$ tetrahedron and three  SiO$_4$ tetrahedra, while
the danburite-like ring includes two BO$_4$ tetrahedra and two SiO$_4$
tetrahedra. See Ref. \cite{Manara20092528} for an illustration of the two
superstructures.) In practice, this revision of the YBD model modifies
the boundary composition $R^*$ to $R^*=0.5+K/2N$, with a value of $N$
between 5 and 6.

In a previous paper, subsequently referred to as
Part~I~\cite{Pedesseau-nbs1}, we have studied the properties of an
NBS system of composition 3Na$_2$O-B$_2$O$_3$-6SiO$_2$ in the {\it
liquid} state.  In the present paper we investigate the properties of the
corresponding {\it glass}. In particular we will focus on the structures
around the boron atoms and how these are embedded into the network. In
addition we will also discuss the electronic and vibrational properties
of the system and how these are related to the structural units. These
studies will allow to gain direct insight on the structural properties
of NBS glass considered and also help to interpret experimental data
for glasses that have a similar composition.

The remaining of the paper is organized as follows: We will start with a
brief reminder of the simulation procedure and give the details on how the
glass samples have been produced. In section \ref{sec:structure} we will
discuss the structural properties of the glass. Section~\ref{sec:vibpropr}
is devoted to the vibrational properties of the system. This is followed
by Sec.~\ref{sec:electronic} in which we present the electronic properties
of the sample as well as the charges of the various species and how these
are related to the local geometry. Section~\ref{sec:dielectric} deals
with the dielectric properties  of the system and in Sec.~\ref{sec:conclu}
we summarize and conclude the paper.

 \section{\label{sec:method}Simulation details}

We have used the Vienna {\it ab initio} package (VASP)
\cite{vasp_code01,vasp_code02} to carry out first principles
molecular dynamics (MD) simulations. The system considered is
3Na$_2$O-B$_2$O$_3$-6SiO$_2$ and the samples had 320 atoms (60 silicon,
180 oxygen, 60 sodium and 20 boron atoms). The size of the cubic box
was 15.97 \AA, which corresponds to a mass density of 2.51 g/cm$^3$ 
(the experimental value of the density for NBS \cite{OMazurin_book})
and periodic boundary conditions were used. The electronic structure was
treated through the Kohn-Sham (KS) formulation of the Density Functional
Theory (DFT)  \cite{Kohn1965,RMartin_book}  using the generalized gradient
approximation (GGA) and the PBEsol functional \cite{GGAPBE,PBEsol}. The
KS orbitals were expanded in a plane wave basis at the $\Gamma$-point
of the supercell of the systems, and  the electron-ion interaction was
described within the projector-augmented-wave formalism (PAW) \cite{PAW}.
The choice of the functional as well as other simulations parameters
have been detailed in Ref.~\cite{Pedesseau-nbs1}.

Since the properties of the glass will depend on their history we
have considered different protocols to generate  our glass samples
(more details are given in Ref.~\cite{Pedesseau-nbs1}): Two independent
samples were equilibrated at 3000~K, i.e. in the liquid. We then started a
linear quench first from 3000~K to 2000~K (using a quench rate of $2\times
10^{14}$ K.s$^{-1}$) and then a quench from 2000~K to 300~K with a rate
of $1.7\times 10^{15}$ K.s$^{-1}$. Furthermore we used four independent
liquid configurations at 2200~K and started from them a quench to 1200~K
with a rate of $2\times 10^{14}$ K.s$^{-1}$, followed by a quench with
rate $9\times 10^{14}$ K.s$^{-1}$ to 300~K.

Four of these six glass samples were annealed for 2~ps keeping the
temperature fixed at 300~K, and for two of them the same annealing
was done for 15~ps. Despite the different quench and annealing history
of the six samples, we did not detect any significant dependence of
their properties on the production process. In the following we will
therefore present only the averaged results.

For the electronic densities of states and vibrational properties,
we quenched the samples to 0~K and relaxed them. This relaxation was
stopped once the  $x,\, y,\, z -$ components of the forces acting
on each atom were less than 10$^{-3}$ eV/\AA. The data discussed in
Secs. \ref{sec:vibpropr} and \ref{sec:dielectric} were obtained by
averaging over 8 samples, as 2 supplementary samples have been generated
by infinitely fast quenches (i.e. steepest descent) from the two liquid
trajectories at 3700~K.

\section{\label{sec:structure} Structure}

In this section we will discuss the structural properties of the glass,
notably the environment of three and four-fold coordinated boron atoms
as well as the arrangement of the sodium atoms around the network formers.

\subsection{\label{sec:PDFglass}Radial pair distribution functions}

The partial radial pair distribution function (PDF) $g_{\alpha\beta}(r)$
is proportional to the probability to find an atom of type $\alpha$
at a distance $r$ from an atom of type $\beta$. The various PDFs have
been shown in Figs.~1 and 2 of Part~I at different temperatures for
the liquid as well as for the glass state~\cite{Pedesseau-nbs1}. For the latter we have
found that the PDFs for Si-O and B-O go to zero between the first and
second nearest neighbor peak.  This feature allows thus to define in
a unambiguous manner which oxygen atom is a first nearest neighbor of
a given Si or B atom and hence to determine the number of  O atoms that
are bonded to the network formers. From that information we then can
identify the so-called $Q^{(n)}-$species, i.e.~the Si atoms that have
four oxygen atoms as nearest neighbor, $n$ of which are bonded to another
network forming atom. In Table~\ref{table-structure} we summarize the
so obtained information regarding the various length of the bonds, the
fraction of $Q^{(n)}-$species, as well as the concentration of three and
four fold coordinated boron atoms. The same quantities are also given
for 2200~K, i.e. the liquid at the lowest temperature at which we have
been able to equilibrate the system. As expected, the values of these
quantities in the liquid are quite close to the ones in the glass state,
since, due to the rapid quenching, the main difference in the structure
is an anharmonic relaxation of the positions of the atoms.

\begin{table}
\centering
 \begin{tabular}{c|c|c}
 &  \quad  glass \quad & \quad  liquid at 2200 K\quad \\
 \hline
Si-O bond length (in \AA ) &  1.64  &  1.63  \\
Si-BO bond length  &  1.64 &  1.63 \\
Si-NBO bond length & 1.58   &  1.58\\
B-O bond length &  1.42  &  1.38 \\
B-BO bond length &  1.42  &  1.40 \\
B-NBO bond length & 1.33  &  1.30  \\
Na-O bond length &  2.29  &  2.24 \\
$\widehat{\mathrm{SiOSi}}\, (^{\mathrm o})$  bond angle  & 132.7 (15.0) & 130.9 (18.1)\\
$^{[5]}$Si (\%)&  7.5 &  10.6 \\
 \Q 4  (\%)&  35.3 & 33.6 \\
  \Q 3  (\%) & 49.8   & 48.2 \\
  \Q 2 (\%) & 6.8  &  7.0\\
 \Q 1 (\%) & 0.3  &  0.1\\
 BO (\%) &  73.2     &  71.9\\
 NBO (\%) &  26.7    &  27.3\\
 TBO (\%) &   0.1   & 0.8 \\
$^{[4]}$B (\%)& 37.0 &   30.5\\
$^{[3]}$B (\%)&  63.0 &  67.8 \\
\hline
 \end{tabular}
\caption{Some structural features  of the NBS glass and as well as
of the liquid at 2200K: Bond lengths, bond angles, concentration
of $Q^{(i)}$ units, oxygen species as well as boron coordinations.
For the bond lengths, the values correspond to the first peak
positions of the corresponding PDFs plotted in Figs.~1 and 2 of
Ref.~\protect\cite{Pedesseau-nbs1}. For the bond angle we also give in parenthesis 
the standard deviation of the distribution. 
}
\label{table-structure}
\end{table}

\begin{figure}
\includegraphics[width=0.43\textwidth]{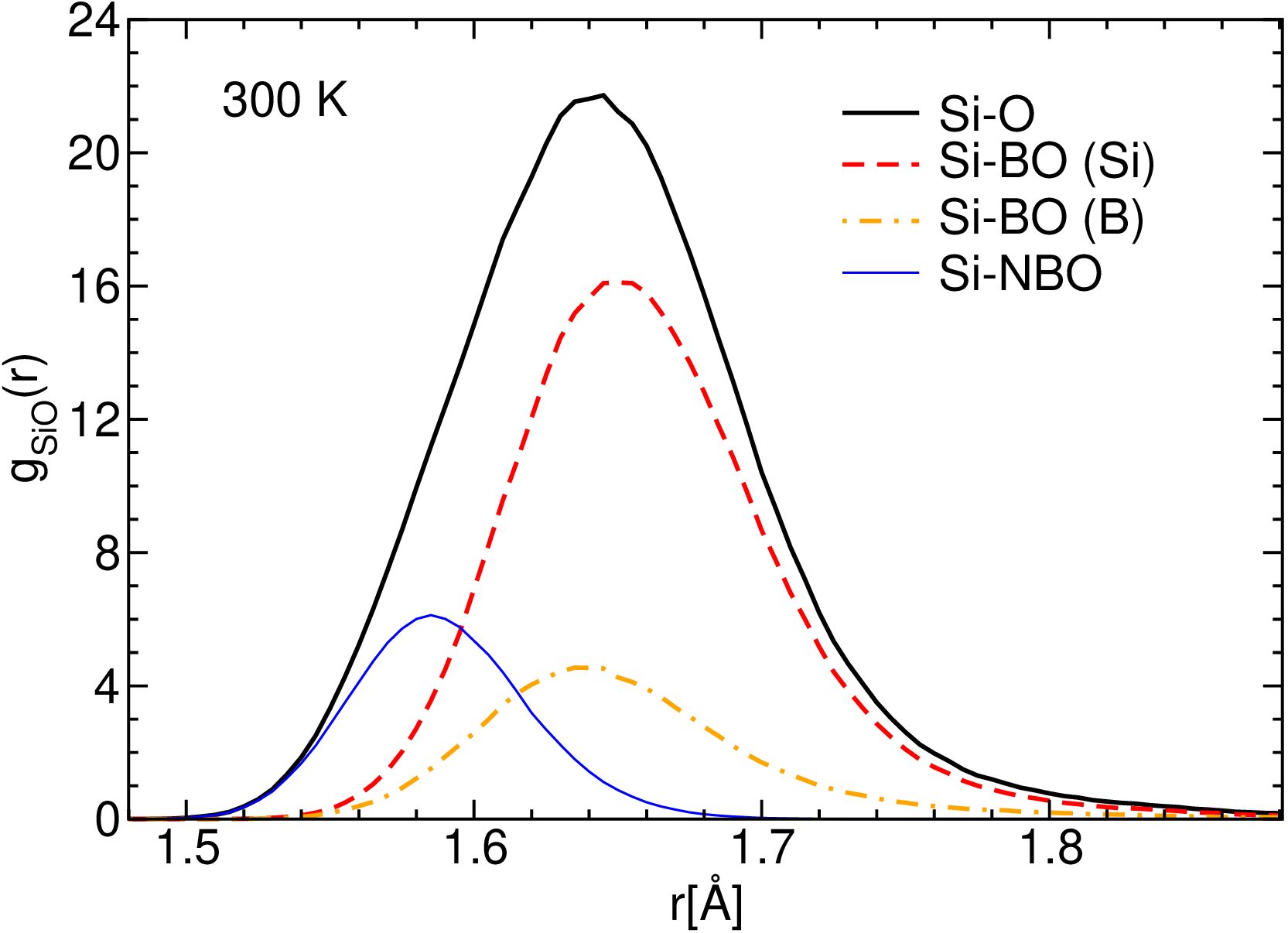}
 \caption{\label{fig:fig1-grSiO}
First peak of the Si-O PDF for the glass decomposed by taking into account
 the oxygen type (BO and NBO) as well as the nature of the type of the second
neighbor of the O atom. }
\end{figure}

From Fig.~1a of Part~I (Inset) one sees that the first nearest neighbor
peak in $g_{\rm SiO}(r)$ is rather sharp. However, a closer inspection
shows that the peak is in reality a superposition of three distinct
peaks that are at slightly different positions. This is documented
in Fig.~\ref{fig:fig1-grSiO}  where we show this PDF decomposed into
contributions according to the different oxygen species: Bridging oxygen
(BO) linking to either Si or B and non-bridging oxygen (NBO). We see that
the Si-O-Si-peak is at a distance ($r_{\rm max}\approx 1.65$~\AA) that
is only slightly larger than the one for Si-O-B ($r_{\rm max}\approx
1.64$~\AA), which shows that in this case the nature of the second
nearest neighbor does not influence the bond distance to the first
nearest neighbor.  More important is the observation that the peak
in Si-O-Si is significantly higher than the one for Si-O-B. If the
network would be completely random, one would expect a factor of three
(since we have 3 times more Si atoms than B atoms). However, we find
that the ratio between the peak height is significantly higher than
3, i.e. the mixing of the two networks  is not ideal, in qualitative
agreement with the observation in Part I, where we found that the Si
and B atoms undergo a microphase separation (see the partial structure
factor $S_{\rm SiB}(q)$ shown in Fig.~6b of Part~I). Finally we see
that the distribution for the NBO shown in Fig.~\ref{fig:fig1-grSiO}
indicates that this oxygen species is relatively close to the Si atom
($r_{\rm max}\approx 1.58$~\AA), a result that is reasonable in terms
of charge balance and in good qualitative agreement with previous
\ab simulations for binary sodium and lithium silicate glasses
\cite{Ispas2001,Donadio2004214205,Du2006114702,Ispas2010} as well as
more complex bioactive glasses \cite{Tilocca2008084504}.


\begin{figure}
 \includegraphics[width=0.43\textwidth]{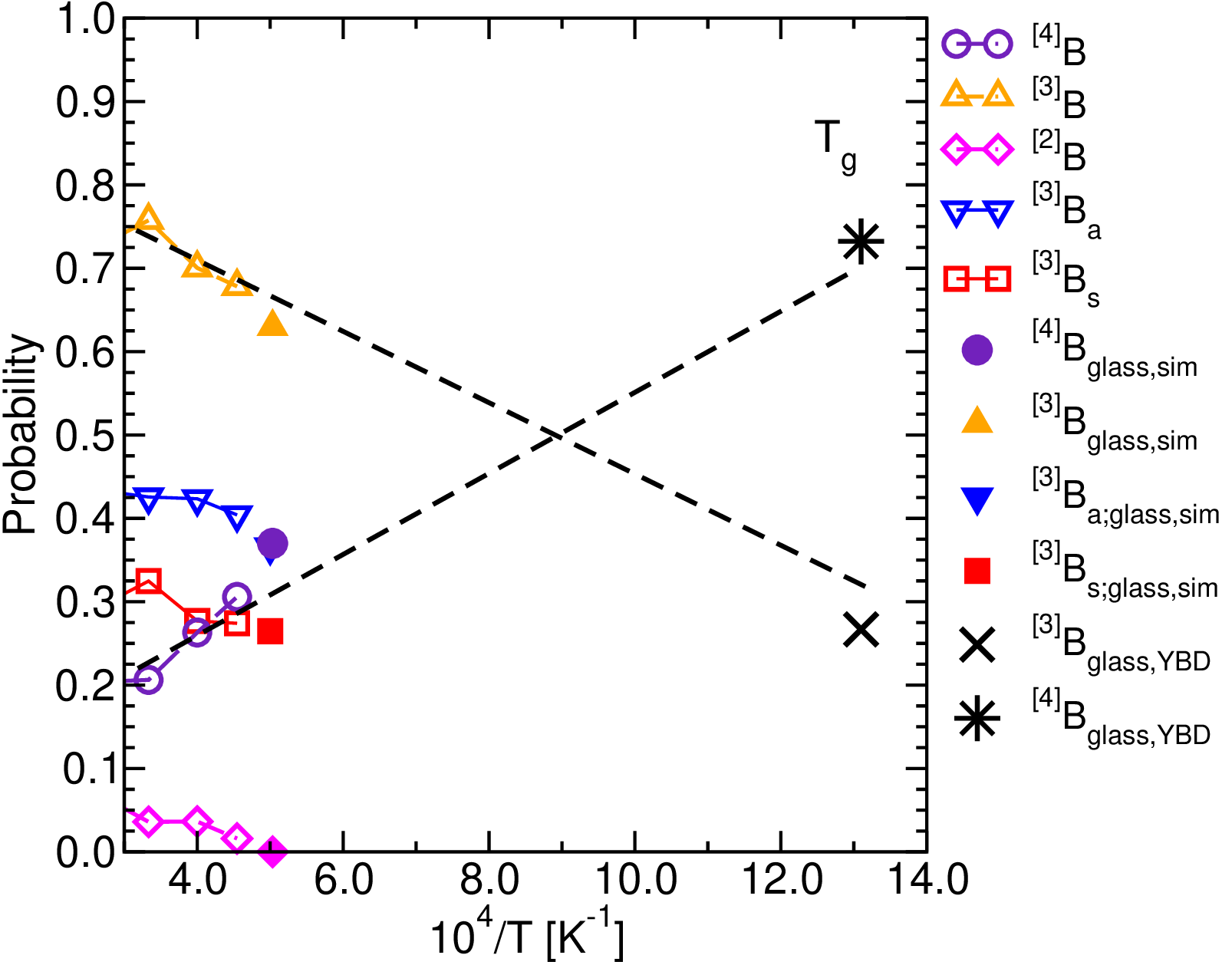}
 \caption{\label{fig:fig2-b3b4-to-Tg}
Temperature dependence in the liquid state of the concentrations of 2-,
3- and 4-fold coordinated borons as well as the ones of the symmetric and
asymmetric 3-fold borons, $^{[3]}\mathrm B_s$ and $^{[3]}\mathrm B_a$
respectively. The filled symbols are the corresponding concentrations
in the glass generated by the fast quench in the simulation. The dashed
lines are the extrapolation of the 3- and 4-fold coordinated borons to low
temperatures taking into account that they have to sum up to 100\%. The
star and cross are the corresponding concentrations as predicted by the YBD model.}
\end{figure}


 \begin{figure}
 \includegraphics[width=0.43\textwidth]{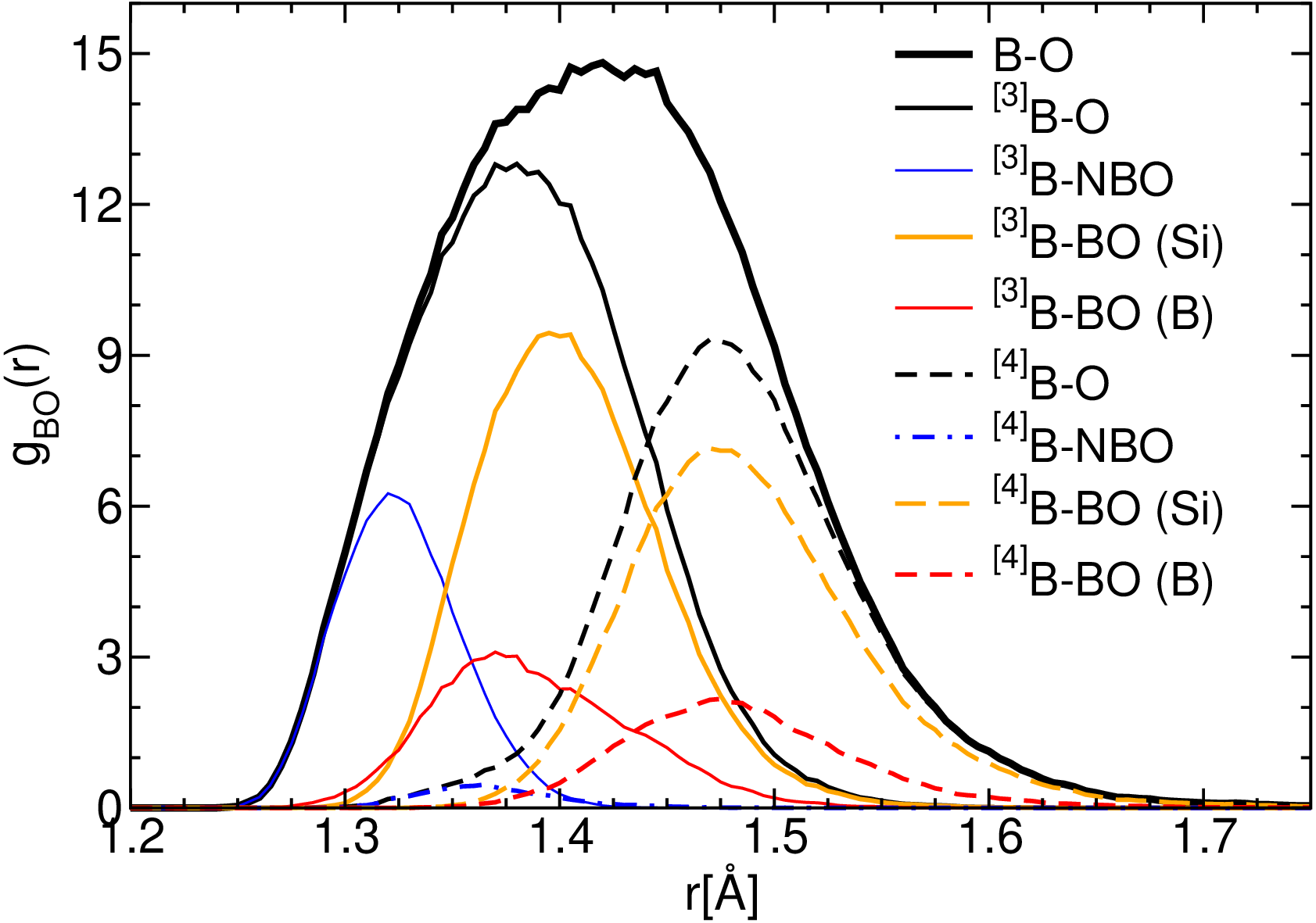}
 \caption{\label{fig:fig3-grBO}
First peak of the B-O PDF for the  glass, decomposed into
contributions from $^{[3]}$B and $^{[4]}$B, into oxygen's speciation BO
and NBO,  as well as according the second network-former of the bridge
B-O-T, for T=Si, B. }
\end{figure}


\begin{figure}
   \includegraphics[width=0.43\textwidth]{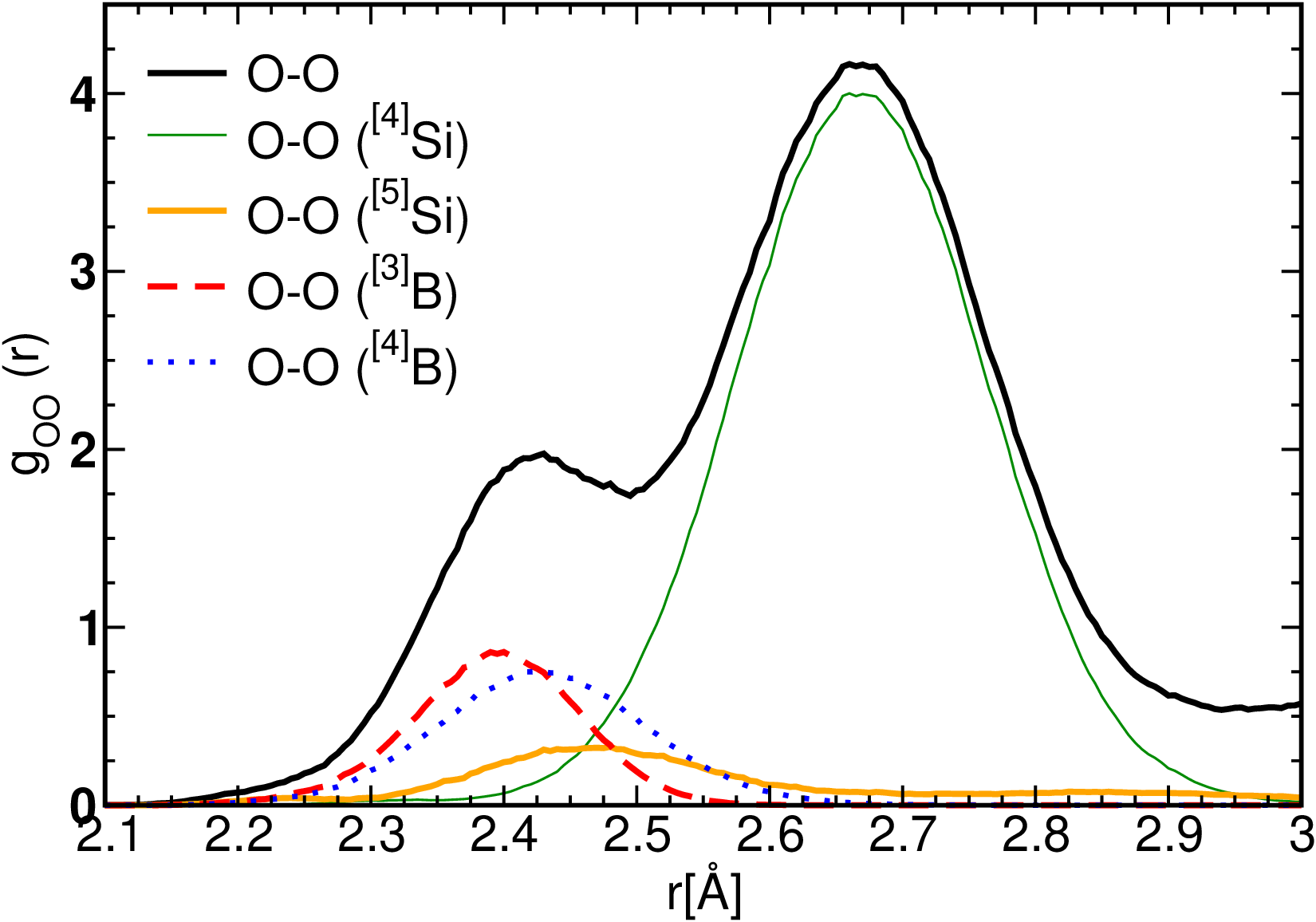}
    \caption{\label{fig:fig4-grOO}
First peak of the O-O PDF decomposed into contributions of oxygen atoms that are associated
with $^{[3]}$B, $^{[4]}$B, and four and five fold coordinated Si.}
\end{figure}


Regarding the PDF for B-O, we will see below that it is useful to
decompose this correlation function into contributions that stem from
the trigonal motifs $^{[3]}$B and the tetragonal ones, i.e. $^{[4]}$B,
since these are important building blocks in the network. Before doing
this it it, however, important to see to what extent the concentration
of these motifs depend on the way we have produced the glass. Although
the structure of all glasses will depend to some extent on the
production history, this dependence is usually mild and hence can,
to a first approximation, be neglected~\cite{Vollmayr1996}. However, the
case of borate systems are special since in the liquid state there is
a significant conversion of three-fold coordinated boron atoms into
four-fold coordinated ones if temperature is decreased and hence it
must be expected that the properties of the glass depend strongly
on its fictive temperature. To estimate this dependence we show in
Fig.~\ref{fig:fig2-b3b4-to-Tg} the $T-$dependence of the concentration of
2, 3 and 4-fold coordinated boron atoms in the liquid (open symbols). As
can be seen, the concentration of 2-fold coordinated boron atoms is
small, decreases very rapidly, and vanishes at around 2000~K. Hence
we conclude that this local structure is not relevant in this type
of glass, in agreement with expectation and experiments. Also the
concentration of $^{[3]}$B decreases if $T$ is lowered, but remains
at relatively high values in the whole temperature range at which we
can probe the equilibrium properties of the liquid. In contrast to
this the concentration of $^{[4]}$B is relatively small at high $T$,
but increases steadily if temperature is lowered. The significant
$T-$dependence that we find for the concentration of $^{[3]}$B
and $^{[4]}$B is in qualitative agreement with results from NMR
experiments in which it has been found for various borosilicate 
glasses that an increase of the quench rate by 4-6 orders of magnitude
leads to a decrease of the concentration of $^{[4]}$B by about
5-7\%~\cite{Sen199829,Sen199984,Kiczenski20053571,Angeli2012054110},
that the concentration of $^{[3]}$B is increasing with increasing
$T$~\cite{Sen199984,Majerus2003024210,Wu20113944,Angeli2012054110}, or
with studies on probes at high temperatures in which it has been shown
that an increase of annealing time of the glass samples leads to higher
concentration of $^{[4]}$B~\cite{Sen2007094203}.

In view of this quite strong $T-$dependence of the concentrations of
$^{[3]}$B and $^{[4]}$B, it is evident that a glass produced with the
high quenching rates imposed by the {\it ab initio} simulations will
not give a reliable value for these concentrations in a real glass.
This is seen in Fig.~\ref{fig:fig2-b3b4-to-Tg} where we also show these
concentrations for the glass we have obtained (filled symbols). As
expected these concentrations are very close to the values we have
extracted for these quantities in the liquid at the lowest temperature,
i.e. 63\% and 37\% for $^{[3]}$B and $^{[4]}$B, respectively. However,
since we have determined the $T-$dependence of these concentrations,
we can extrapolate them to the temperature at which the {\it real} liquid
undergoes a glass transition temperature and hence obtain a more reliable
prediction for these values. This extrapolation, in which we have taken
into account that the sum of the two concentrations must add up to
100\%, is included in Fig.~\ref{fig:fig2-b3b4-to-Tg} as well (dashed
lines). We see that for a $T_g \approx 760$K one obtains values that
are close to the ones predicted by the YBD model (25\% and 75\% for
$^{[3]}$B and $^{[4]}$B, respectively, show in the figure as well)
and also to the ones estimated from XANES experiments (22\% and 78\%,
respectively)~\cite{Fleet1999233}. Hence we can conclude that our
simulations do indeed allow to predict the mentioned concentrations in
the real glass, {\it if one takes into account their $T-$dependence}.

Having understood the $T-$dependence of the concentrations of $^{[3]}$B
and $^{[4]}$B, we return to the discussion of the PDF for B-O and in
Fig.~\ref{fig:fig3-grBO} we show the first peak of this function.
In view of the importance of the local motifs $^{[3]}$B and $^{[4]}$B
we have decomposed this peak into the contributions from trigonal and
tetrahedral boron (full and dashed lines, respectively).

From the graph we see that the length of the B-O bond is 1.38 \AA~and
1.47~\AA~for $^{[3]}$B-O and $^{[4]}$B-O, respectively. These values are
in excellent agreement with the ones obtained from neutron scattering
experiments on \boric\, and binary borate glasses which gave 1.37~\AA~and
1.47~\AA, respectively~\cite{Swenson19959310} or on sodium diborate
glass which gave 1.38 \AA\, for the $^{[3]}$B-O distance and 1.485 \AA\,
for the $^{[4]}$B-O distance~\cite{Majerus2003024210}. Our values are
also close to the ones obtained within a classical MD simulation of
borosilicate glasses \cite{Kieu20113313,Inoue201212325}, thus showing
that these distances are basically independent of the composition. It is
therefore noteworthy that a reverse Monte Carlo simulation of borosilicate
glasses with quite similar Na$_2$O/\boric\, and \silica/\boric\, ratios,
but containing 5 \% BaO  - 5\%Zr$_2$O gave a $^{[4]}$B-O distance around
1.60~\AA~\cite{Fabian20072084}, which indicates that this approach does
not give reliable results.

Although the position of the first peaks in $^{[3]}$B-O and $^{[4]}$B-O
are very close to the ones found in pure \boric\, or binary borates,
one has to realize that the local structure of the present system is
more complex. In particular the presence of Na makes that there is a
significant concentration of NBO, with the consequence that the B-O
distance shows a relatively wide distribution. This can be seen in
Fig.~\ref{fig:fig3-grBO} where we show also the decomposition of the
$^{[3]}$B-O and $^{[4]}$B-O distributions into the parts involving
bridging and non-bridging oxygen atoms. From the size of the peaks one
sees immediately that the NBO play only a minor role for the $^{[4]}$B
environment (1.2\%) whereas in the case of the $^{[3]}$B structures
they account 18\% of the bonds.  Hence we conclude that NBOs are mainly
associated to the trigonal motifs.

The location of the NBO peaks is shifted significantly to smaller
distances with respect to the ones for the BO, a result that, in
view of the reduced connectivity of the NBOs, is reasonable. For the
$^{[3]}$B-NBO distribution we find the peak at 1.32~\AA~, whereas the one
for $^{[3]}$B-BO it is at 1.39~\AA. That the {\it mean} position of the
$^{[3]}$B-O peak  is basically the same as the ones found for \boric\,
(for which one does not expect to have NBO) is therefore only due to
the effect of averaging the peaks for the $^{[3]}$B-NBO and $^{[3]}$B-BO.

Finally we mention that the length of the B-O bond can also be influenced
by the second atom that is connected to the oxygen: For the case of a
$^{[3]}$B-O bond in which the O is attached to a B or Si atom the bond
length is 1.37\AA\, and 1.40~\AA, respectively. However, for the case
of an $^{[4]}$B-O bond we do not find such a difference, thus showing
that the boron triangles and tetrahedra have quite different rigidity.

Figures~\ref{fig:fig1-grSiO} and \ref{fig:fig3-grBO} show that there
is a substantial amount of NBOs in our system and we find that the
total concentration is 27~\% (i.e.~the BO have a concentration of
73~\%). Although we have just shown that the $^{[3]}$B triangles contain
a significant fraction of NBOs and that the concentration of the former
depends significantly on $T$, the overall concentration of NBOs does
not depend strongly on temperature (see Fig.~4 in Part I). The reason
for this is that on one hand we have three times more Si than B  atoms,
i.e. the strong $T-$dependences of $^{[3]}$B is not that relevant, and
on the other hand some of the NBOs associated to $^{[3]}$B are probably
accommodated by the Si atoms, thus weakening the $T-$dependence of the
overall NBO concentration. As a result one can make an extrapolation of
the $T-$dependent BO concentration to low temperatures and estimate that
around $T_g$ this value is slightly above 80~\%. This result agrees very
well with the prediction of the YBD model which foresees 80~\% and 20~\%
for BO and NBO concentration, respectively, and also with results from
$^{17}$O NMR experiments~\cite{Wang19991519}.

For the BOs we have found that 45~\% belong to Si-O-Si bridges, 25~\%
to Si-O-B bridges, and only 4~\% to B-O-B bridges. For the present
composition, the YBD model predicts the presence of reedmergnerite, diborate
and pyroborate supra-structural units, and consequently there should be
5\% B-O-B, 28\% Si-O-B and 47\% Si-O-Si, in very good agreement with our
findings. However, in the $^{17}$O NMR experiments of Wang and Stebbins
only a low concentration of Si-O-B has been found~\cite{Wang19991519},
which might imply that the Si and B network are not perfectly mixed (see
also \cite{Stebbins199880,Sen199829}). Although such an non-ideal mixing
behavior is indeed compatible with our simulation data (we recall that
in Part I, Fig.~6b, we have shown that at small wave-vectors the partial
structure factor for Si-B does not go to zero), the effect is so small
that it does not affect the concentration of the Si-O-B connections and
hence this value can still be relatively elevated.

Finally we mention that the possible connection of NBO to the
$^{[3]}$B triangles makes that the latter can be separated into two
classes~\cite{Dell19831}: The ones including one or two NBOs, called
asymmetric units (labelled hereafter $^{[3]}$B$_a$), and the ones
completely connected to the network through bridging oxygens, i.e. with
zero NBOs, called symmetric units (labelled hereafter $^{[3]}$B$_s$).
The YBD model does predict the relative concentration of these two units
and for the current composition one should have 55\% of $^{[3]}$B$_a$
and 45\% of $^{[3]}$B$_s$. The corresponding values we have measured for
our glasses are around 58\% for $^{[3]}$B$_a$ and 42\% for $^{[3]}$B$_s$,
thus they agree very well with the predictions of the YBD model, although
our error bars for these concentrations are relatively large.  In view
that one has a strong cooling rate dependence of the concentrations for
$^{[3]}$B and $^{[4]}$B, this agreement might surprise.  However, NMR
experiments \cite{Sen199829} have probed the cooling rate dependences
for these two $^{[3]}$B structures and found that the concentration of
$^{[3]}$B$_s$ is almost independent of this rate, whereas the ones of the
asymmetric units shows a decrease of about  6\% when the cooling rate
was decreased over four orders of magnitude. Thus we can conclude that
the transformation of $^{[3]}$B into $^{[4]}$B units is mainly made by
eliminating the $^{[3]}$B$_a$ structures. Unfortunately the temperature
dependence of  $^{[3]}$B$_s$  and $^{[3]}$B$_a$  in the liquid state
(shown in Fig.~\ref{fig:fig2-b3b4-to-Tg}) do not allow us to confirm
this conclusion since the large error bars are too large.

In Fig.~1d of Part I we have seen that the first nearest neighbor peak
in $g_{\rm OO}(r)$ is split into two. The reason for this feature is given in
Fig.~\ref{fig:fig4-grOO} where we have decomposed this correlation
function into contributions from first-neighbor O-O distances in
SiO$_4$, BO$_3$, BO$_4$, and SiO$_5$ units, the latter occurring with
a probability of around 7~\%. From the graph we recognize that the
peak at 2.40~\AA\, is due to the O-O distance within the trigonal and
tetrahedral boron structures  which give rise to peaks at 2.38 \AA\, and
2.42 \AA, respectively. These values agree well with the ones obtained
from neutron scattering experiments on \boric\, glass and on binary
borate glasses containing modifiers for which, depending of the nature
and proportion of the modifier, distances ranging from 2.38 to 2.42 \AA\,
are found~\cite{Swenson19959310}. For trigonal units, the present O-O
distance agrees also with that extracted from classical MD simulations
for \boric\, glass, which was 2.37~\AA~\cite{Takada19958693}.

The main nearest neighbor peak in $g_{\mathrm{OO}}(r)$ is found
at  2.66~\AA\, and its presence is entirely due to the SiO$_4$
tetrahedra. This distance is close to the one found in pure SiO$_2$
glass~\cite{Giacomazzi2009064202}, but this does not imply that the
geometry of this structural entity is not affected by the presence
of other network-formers and/or modifiers.  This coincidence
is only due to the effect of  averaging over distorted SiO$_4$
tetrahedra with short Si-NBO bonds and long Si-BO bonds (see
Fig.~\ref{fig:fig1-grSiO} and Tab.~\ref{table-structure}), as already
reported in previous {\it ab initio} simulations  for other silicates
\cite{Ispas2001,Benoit2001,Du2006114702,Ispas2010,Tilocca2010014701}.
Finally we mention that the O-O distance corresponding to the  SiO$_5$
units is somewhat smaller than the one for the  SiO$_4$ tetrahedra.
Since only 2.5\% of the defective SiO$_5$ structures are associated with
a NBO, the reduced Si-O distance is not related to these non-bridging
oxygen atoms.

From Figs. 2a, b and c in Part I we see that also the PDFs for Si-Si, Si-B
and B-B show a split first nearest neighbor peak. A detailed inspection of
these features shows that they are related to the presence of defective
edge-sharing units and/or defective SiO$_5$ units which, due to the high
quenching rate, are more abundant in our samples than in real glasses.

This is also the case for the first peak in the B-B correlation: We
have found it to be present only in one of the samples and therefore
consider it to be atypical. The relevant peak in this PDF is located at
2.60 \AA\, which is in very good agreement with recent neutron scattering
studies by Michel {\it et al.}~\cite{Michel2013169}  who found for the same
composition the distance 2.64~\AA\, and {\it ab initio} simulations of
\boric\, for which one finds 2.55\AA~\cite{Ohmura2008224206}.  However, other
neutron scattering experiments have found that the peak is located
between 2.77 and 2.85~\AA\, for \boric\, and alkali borate glasses with
comparable alkali content, respectively~\cite{Swenson19959310}. This
shows that extracting such detailed information from neutron scattering
experiments is not a trivial task. 

We also point out that this distance might be affected by the high
quench rate used in computer simulations. Since we have seen that our
simulations predict a concentration of $^{[3]}$B that is too high, it
can be expected that a smaller cooling rate (which will give rise to
more $^{[4]}$B units) will lead to a B-B distance that is larger, since
we have seen in Fig.~\ref{fig:fig3-grBO} that the distance $^{[3]}$B-O
is smaller than the $^{[4]}$B-O one.  Finally we mention that this too
low concentration of $^{[4]}$B hinders us to check the validity of the
so-called ``tetrahedral boron avoidance'' principle, i.e. the trend
seen in NMR experiments that the $^{[4]}$B units have the tendency of
not being nearest neighbor~\cite{Du200310063,Du2004196}.

\begin{figure}
   \includegraphics[width=0.43\textwidth]{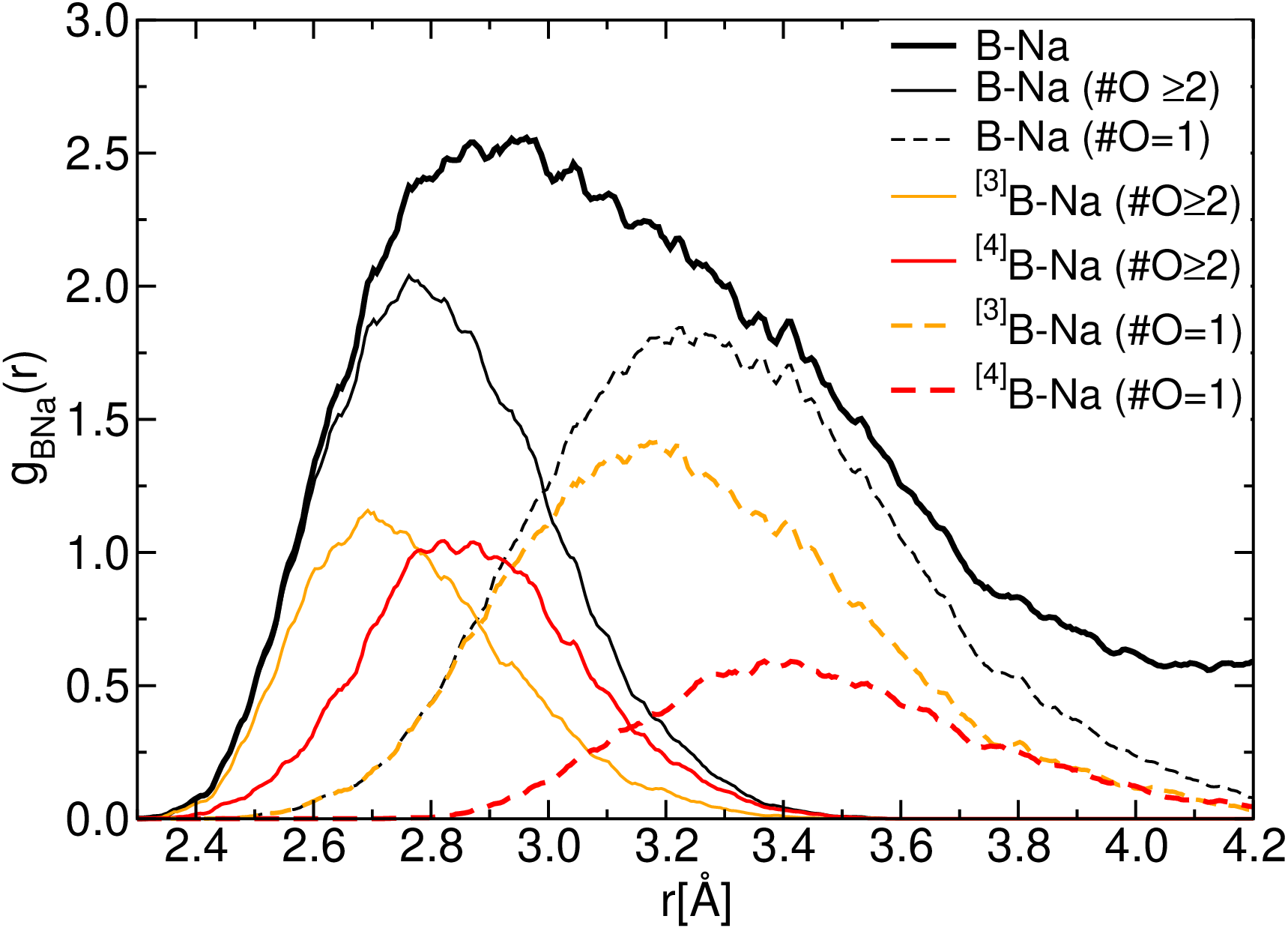}
    \caption{\label{fig:fig5-grBNa}
Decomposition of the 1st peak of the
B-Na PDF for the  glass according to boron coordination as well
as to the number of oxygens that are shared between the B and Na atom.}
\end{figure}

To understand how the alkali atoms arrange around the BO$_3$ and BO$_4$
units we show in Fig.~\ref{fig:fig5-grBNa} the corresponding PDF. One
clearly sees that the nearest neighbor peak (bold solid line) is the sum
of two contributions. We have found that it is reasonable to decompose the
main peak into B-Na pairs that share exactly one oxygen atom (thin solid
line) or more than one O atom (thin dashed line).  Thus we recognize
that the shoulder at around 3.25 \AA\, is due to the B-Na pairs that
share one oxygen atom, whereas the main peak, located at  2.77 \AA,
is produced by the pairs that share more than one oxygen. That sharing
more than one oxygen leads to a shorter distance between the B-Na pairs
is reasonable, since these oxygens give rise to an effective attraction
between the cations. The two mentioned peaks can be decomposed further
into the contributions stemming from $^{[3]}$B and $^{[4]}$B. We find,
see Fig.~\ref{fig:fig5-grBNa}, that the $^{[3]}$B-Na distances are shorter than the one
in $^{[4]}$B-Na, which is reasonable since the tetrahedral structures
are larger than the trigonal ones. Furthermore we recognize from this
decomposition that the probability that a $^{[4]}$B shares exactly one
oxygen atom with a Na is significantly smaller than the corresponding
probability for a $^{[3]}$B. This result is related to the three
dimensional distribution of the Na atoms around the boron units, as we
will show next.

\begin{figure}
   \includegraphics[width=0.43\textwidth]{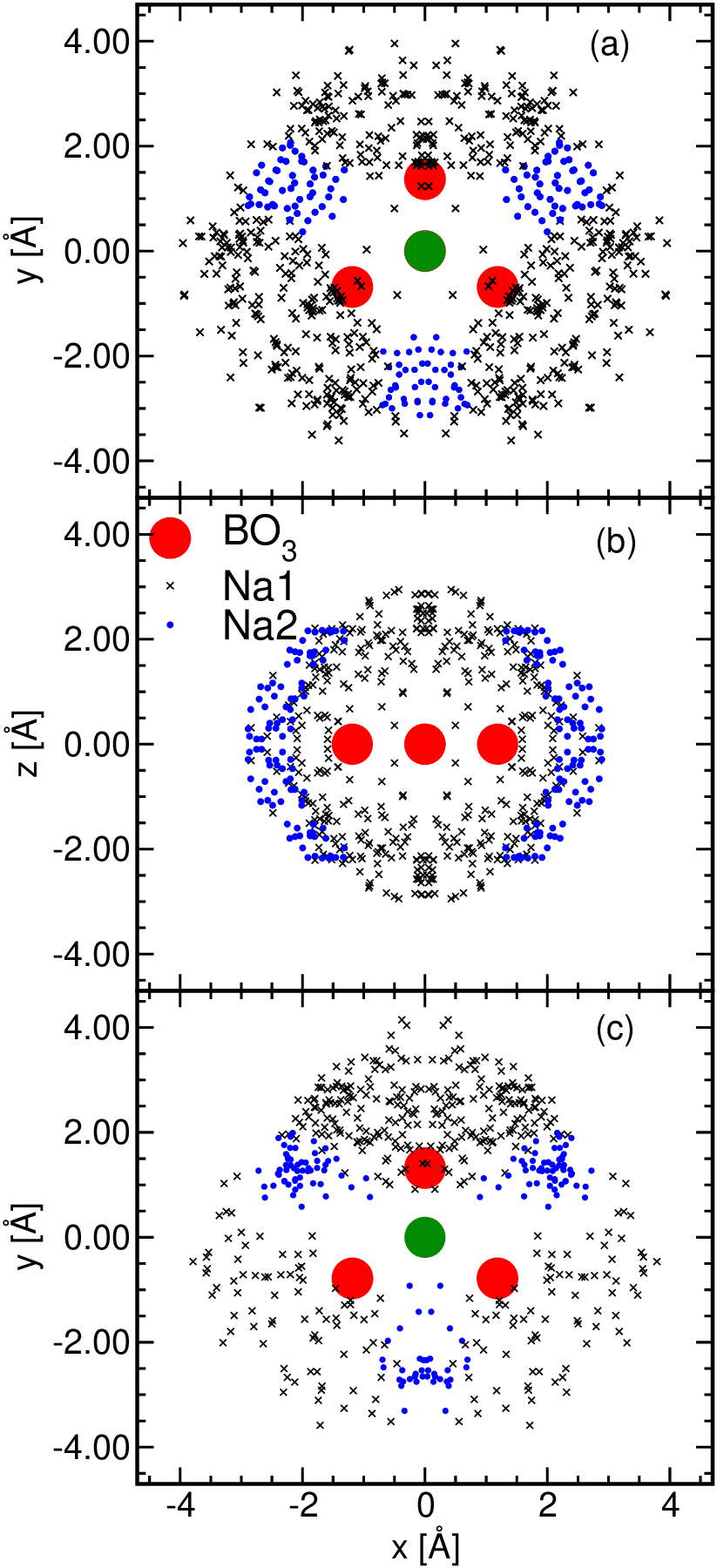}
    \caption{\label{fig:fig6-b3na} Distribution of the Na atoms around a
BO$_3$ unit. The crosses and circles show Na atoms that share, respectively,
exactly one and two oxygen atom(s) with the central B atoms.  a) BO$_3$
in top view (The oxygen on the top is oxygen \#1). b) BO$_3$ in side view. Only Na atoms that are nearest
neighbor to the oxygen atom \#1 (at the center) are shown. c) The asymmetric BO$_3$ unit with two BO and
one NBO (oxygen \#1, on top).}
\end{figure}

Since the PDFs give only information on the relative arrangement of two
atoms it is instructive to consider also three dimensional distribution
functions. In particular we are interested to understand how the Na atoms
are arranged around  the $^{[3]}$B and $^{[4]}$B structures. Since each
$^{[3]}$B triangle or $^{[4]}$B tetrahedron will have a slightly different
geometry we have mapped each of them on an average structure. For
the $^{[3]}$B we thus proceeded as follows: We translated each BO$_3$
triangle, together with its nearest Na, to a coordinate system in which
the B was at the origin and the three O atoms define the $x-y$ plane. In
this plane we set up an equilateral triangle with the vertices at a distance
1.37~\AA\, from the origin (i.e. the B-O distance found in the PDF)
and with one of them lying on the $y-$axis. This triangle defined our
ideal reference structure. The real $^{[3]}$BO$_3$ triangles were then
rotated in this plane such that the average squared distance of the
real oxygens from the ideal positions (given by the reference structure)
were minimized. (We note here that the four atoms of the $^{[3]}$BO$_3$
structure do basically lie in a plane and hence the reference triangle is
a good approximation.) In order to exploit the symmetry of the trigonal
structure we applied all possible symmetry operations to this triangle,
carrying along the Na atoms. These last operations allow thus to improve
the statistics regarding the regions in which the Na atoms prefer
to be. In Fig.~\ref{fig:fig6-b3na}a we show the projection of this
distribution onto the $x-y$ plane and we see that the Na preferential
region presents the expected three fold symmetry along the three BO
bonds. Note that the center of the triangle is basically devoid of Na
atoms, i.e. they prefer to arrange around the three B-O axes. Note that
in the graph we distinguish the Na atoms that share one (crosses) and two
(circles) oxygen atoms with the central B atom. These latter ones are
found close to the angle bisector O-B-O and their distance from the B
atom is slightly smaller than the former ones.  Thus this explains why
the corresponding PDFs, shown in Fig.~\ref{fig:fig5-grBNa}, have their
maximum at different distances.

In panel b) we show the projection of the distribution in the direction
of the $y-$axis. We show only the Na atoms that are nearest neighbor
with the oxygen atom \#1 and we distinguish again atoms that share one
(circles) and two (crosses) oxygen atoms with the central B. We see
that most of the latter atoms are close to the plane spanned by the
three oxygen atoms and that the distribution of (all) the Na is quite
circular around the axes defined by a B-O bond.

To generate panels a) and b) of Fig.~\ref{fig:fig6-b3na} we have only used
those $^{[3]}$B units which have three {\it bridging} oxygen as neighbors.
There is, however, a substantial fraction of $^{[3]}$B which have one
NBO as nearest neighbor (58\%) and hence can be expected to have an
asymmetric distribution of Na atoms around them. That this is indeed the
case as  shown in Fig.~\ref{fig:fig6-b3na}c where we show these triangles
(oriented such that the NBO points upward). We clearly see that the NBO
has significantly more Na around it (on average 3.1 instead of the 1.3
found around the symmetric oxygen in the $^{[3]}$B triangle). From the
figure we also recognize that this enhanced concentration is accompanied
by a decreased concentration of Na around the BO which makes that the
overall number of Na atoms around a $^{[3]}$B with one NBO is equal to
4.8 which has to be compared with 3.2 atoms around a $^{[3]}$B unit with
only BOs.

To study the distribution around the $^{[4]}$B units we have applied three
rigid rotations to each of the real tetrahedra so that its vertices were
as close as possible to the ones of a regular tetrahedron that had a BO
distance equal to 1.47~\AA\, (i.e. the average bond lengths extracted from
the first peaks of the PDFs for the tetrahedra) with one vertex pointing
in the $z-$direction. In order to exploit the symmetry of the structure
we applied again all possible symmetry operation to this tetrahedron,
carrying along the Na atoms.

\begin{figure}
   \includegraphics[width=0.43\textwidth]{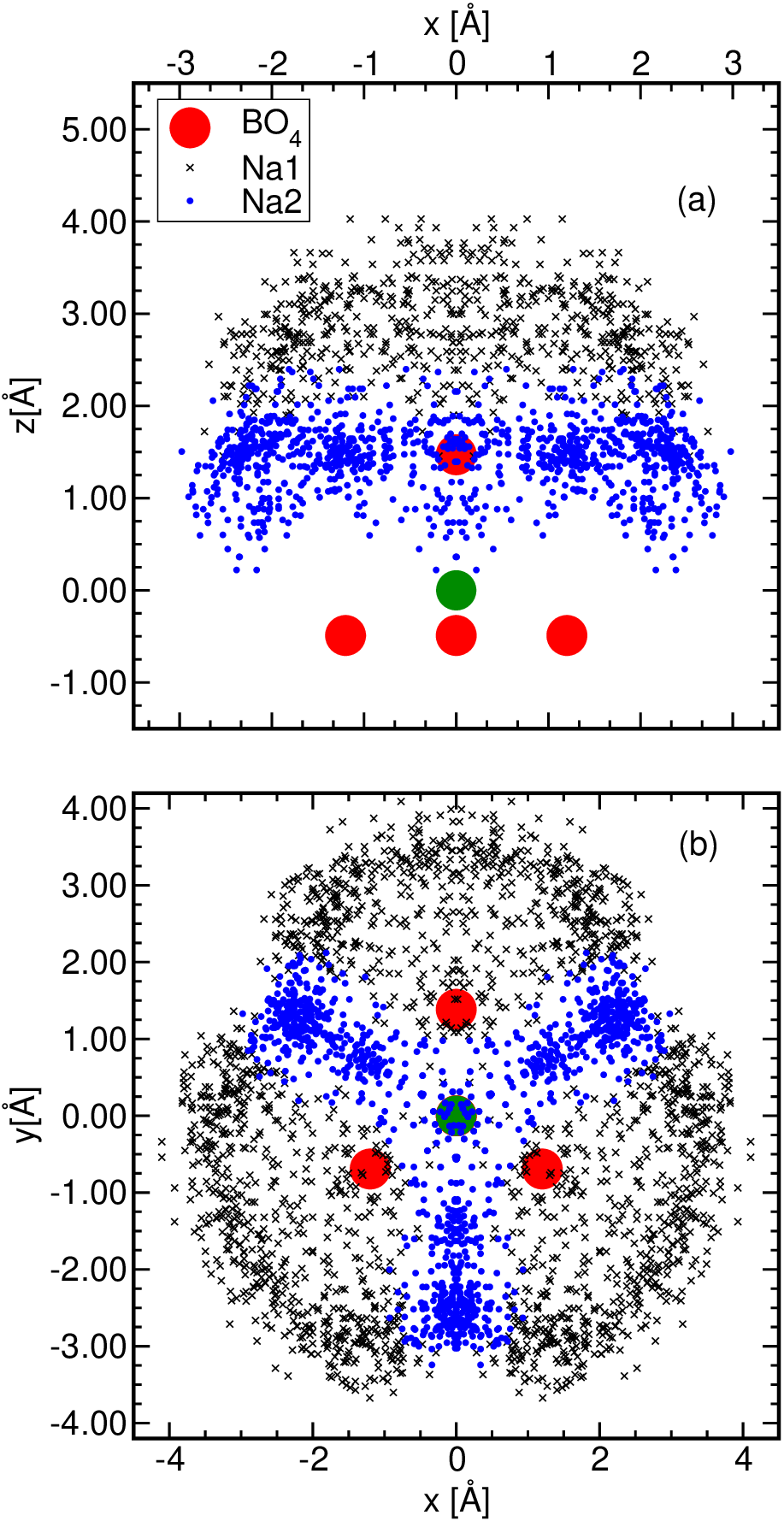}
    \caption{\label{fig:fig7-b4na}
Distribution of the Na atoms around a BO$_4$ tetrahedron. a) Top
view. Only the Na atoms that are first nearest neighbor of oxygen \#1 (the top one) are
shown. b) Bottom view. Only the Na atoms that are not nearest neighbor
of oxygen \#1 (at the center) are shown.}
\end{figure}

In Fig.~\ref{fig:fig7-b4na}a we show the projection of the distribution
onto the $x-z-$plane. For the sake of clarity we show only those Na
atoms that are nearest neighbors of the oxygen atom shown at the top
(e.g. O\#1). We distinguish again between Na atoms that share exactly one
(crosses) or two (circles) oxygen atom(s) with the central B atom. An
inspection of the distribution around the O\#1 as projected on the $x-y-$
plane (not shown) demonstrates that the former Na have a (basically)
circular distribution around the B-O axis whereas the latter Na (circles)
have a distribution with a three-fold symmetry. This symmetry can be
clearly recognized from panel \ref{fig:fig7-b4na}b where we plot the
Na distribution projected onto the $x-y-$plane. (Note that in this
panel the distribution is see from the bottom of the tetrahedron
and that the Na atoms that are first nearest neighbor of oxygen \#1,
located on the positive $z-$axis, are {\it not} shown.)  From the two
projections we also recognize that the Na atoms that share two O with
the central B have typically distances that are smaller than those Na
atoms that share only one, in agreement with the corresponding PDFs
shown in Fig.~\ref{fig:fig5-grBNa}.


 \begin{figure}
   \includegraphics[width=0.43\textwidth]{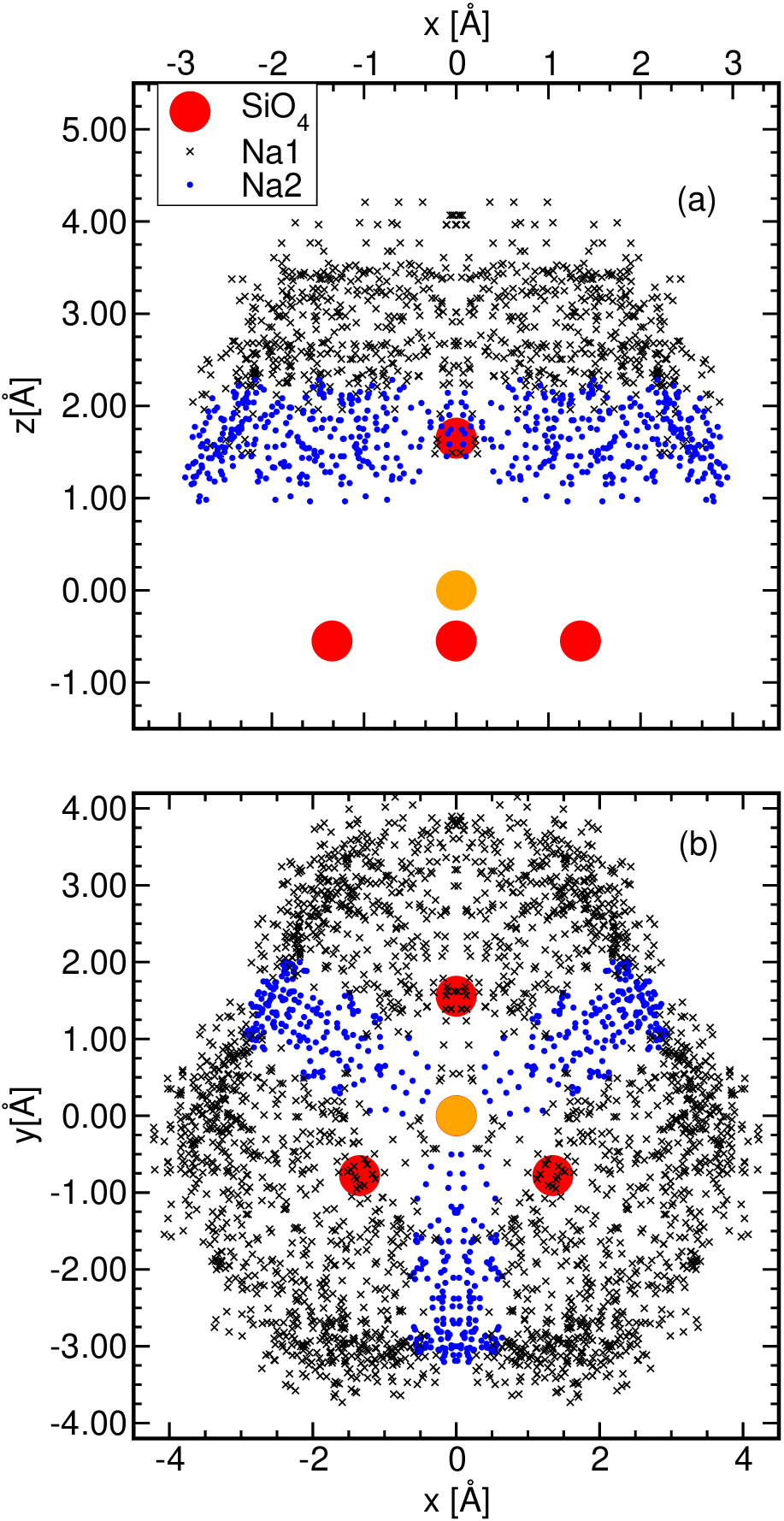}
    \caption{\label{fig:fig8-si4na}
Distribution of the Na atoms around a SiO$_4$ tetrahedron. a) Top
view. Only the Na atoms that are first nearest neighbor of oxygen \#1 are
shown. b) Bottom view. Only the Na atoms that are not nearest neighbor
of oxygen \#1 are shown.}
\end{figure}


Since the BO$_4$ unit have the same symmetry as the SiO$_4$
tetrahedra, it is instructive to compare the Na distribution around
these two structures. For this we show in Fig.~\ref{fig:fig8-si4na}
the distribution for SiO$_4$. (Here we have used the value 1.65 \AA
~for the Si-O distance of the reference tetrahedron and in order to
make the graphs comparable we show in the graphs the same number of Na
as in Fig.~\ref{fig:fig7-b4na}.) The comparison with the corresponding
distribution for BO$_4$ shown in Fig.~\ref{fig:fig7-b4na} reveals that
for the case of SiO$_4$ the Na atoms are concentrated more around the
axis between the cation and the oxygen, thus making that the number of
sodium atoms that share only one oxygen with the central cation (crosses)
is significantly higher. Furthermore we see that the Na which share
two oxygen atoms with the central cation (circles) are spread over a
significantly larger portion of space for the case of BO$_4$. This result
is reasonable since the boron atoms have a smaller effective charge than
the Si atoms, thus restricting the Na atoms less. Hence we see that due
to the effect of charge balance the distribution of Na atoms around the
two tetrahedral structure shows small but significant differences.

\subsection{\label{sec:BADglass}Bond angle distributions in NBS glass}

Complementary and additional information on the local ordering and
organisation of the glass network is obtained from the analysis of the
bond angle distributions (BAD) $P_{\alpha\beta\gamma}(\theta)$ and these
nine distributions are shown in Fig.~9 of Part~I \cite{Pedesseau-nbs1}. As
expected, with decreasing temperature these distributions become
narrower and some of their features more pronounced. At 300~K the
distribution $P_{\mathrm{OSiO}} (\theta)$, Fig.~9a in Part~I, shows
a quite narrow distribution with a peak at about 108$^{\rm o}$, very
close to the ideal intra-tetrahedral O-Si-O angle, and thus similar
to the geometries found for pure \silica\, and  in alkali silicate
glasses~\cite{Ispas2001,Donadio2004214205,Du2006114702,Ispas2010}.
A closer inspection of this distribution reveals the presence of a small
bump at about 90$^{\rm o}$, which is the fingerprint of defective 5-fold
coordinated Si and/or edge-sharing tetrahedra.

\begin{figure}
   \includegraphics[width=0.43\textwidth]{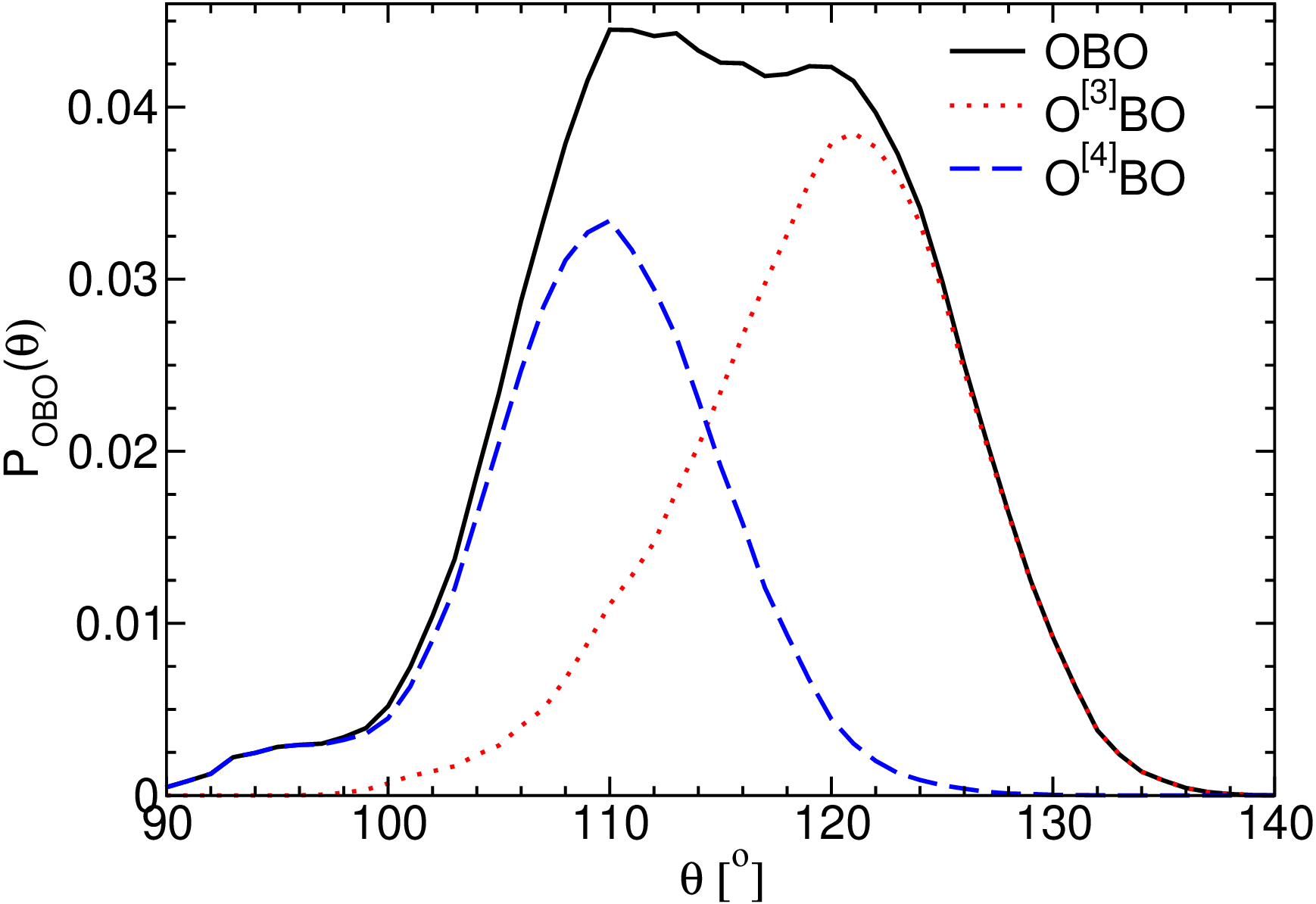}
    \caption{\label{fig:fig9-obo}
Distribution of the bond angle O-B-O and its
decomposition into O$^{[3]}$BO and O$^{[4]}$BO components.}
\end{figure}

The distribution $P_{\mathrm{OBO}} (\theta)$ shown in Fig.~9b of
\cite{Pedesseau-nbs1} is broader than $P_{\mathrm{OSiO}} (\theta)$ and in
fact one sees that it is the sum of two peaks. (The shoulder at around
95$^{\rm o}$ is due to edge sharing triangles/tetrahedra, geometries
that can be expected to vanish if the cooling rate is decreased.) In
Fig.~\ref{fig:fig9-obo} we demonstrate that these peaks are related
to the presence of O-$^{[3]}$B-O and  O-$^{[4]}$B-O, giving rise to
distributions that have a maximum at around 121$^{\rm o}$ and 110$^{\rm
o}$, respectively. These values are close to the expected values for an
 isosceles triangle (120$^{\rm o}$) and an ideal tetrahedron (109$^{\rm
o}$). (We recall that the $^{[3]}$BO$_3$ triangles are basically planar,
similar to the case of pure \boric\, glass~\cite{Umari2005137401}.) In
addition we see that the distribution for the angles O-$^{[3]}$B-O is
asymmetric in that it has a tail to smaller angles. This feature is
related to the presence of $^{[3]}$B that have a NBO, thus making that
the triangular unit has two angles NBO-B-BO at around 123$^{\rm o}$
and a smaller one (BO-B-BO) at around 113$^{\rm o}$.


 \begin{figure}
   \includegraphics[width=0.43\textwidth]{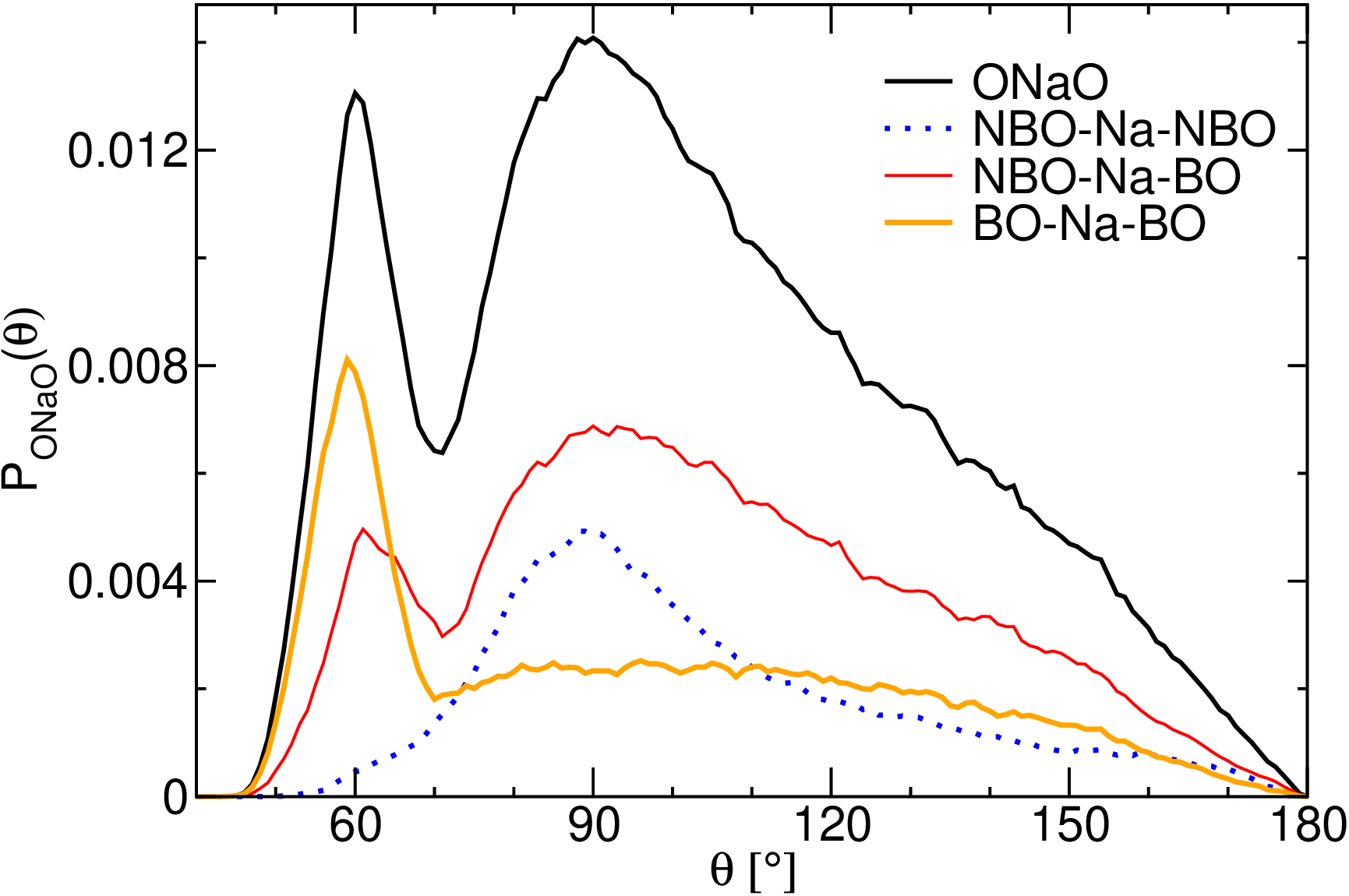}
    \caption{\label{fig:fig10-onao}
Distribution of the bond angle O-Na-O and its decomposition for BO and NBO.}
\end{figure}


Figure \ref{fig:fig10-onao} shows the angle distribution
$P_{\mathrm{ONaO}} (\theta)$.  In Sec. III.C of Part~I
\cite{Pedesseau-nbs1}, we showed that this double peak structure
becomes more pronounced if temperature is decreased (see Fig.~9c
in \cite{Pedesseau-nbs1}). In agreement with previous results from
{\it ab initio} simulations for low-silica alkali-alkaline earth melts
\cite{Tilocca2010014701}, we see that the peak at around 60$^{\mathrm o}$
is related to the BO-Na-BO angle, whereas the one at  90$^{\mathrm o}$ to
the NBO-Na-NBO angle. Finally the distribution for the BO-Na-NBO has peaks
at the two angles.  Since NBO are basically absent in the $^{[4]}$BO$_4$
units, the two latter distributions can thus be associated with Na atoms
that are around Q$^{3}$ and $^{[3]}$BO$_3$ units. Since the concentration
of the $^{[3]}$BO$_3$ units is decreasing with decreasing $T$ and the one
of Q$^{3}$ is increasing (see Fig.~4 in \cite{Pedesseau-nbs1}), it is at
this point difficult to estimate how the distribution $P_{\mathrm{ONaO}}
$ will be in glasses that have been produced with the cooling rates used
in real experiments.

Further insight into the connectivity of the network formers can
be obtained from the BAD centered on the BOs, i.e. the angular
distributions Si-O-Si, B-O-B, and Si-O-B, shown in Part~I, in Figs.~9d,
e, and i, respectively.  All three distributions present small peaks
located at 93$^{\mathrm o}$, 84$^{\mathrm o}$ and 87$^{\mathrm o}$,
respectively, indicating the presence of the defects mentioned in the
previous sub-sections (SiO$_5$ and edge-sharing). Therefore it can
be expected that with decreasing cooling rate these peaks disappear
(see Fig.~5b in Part~I) and are not present in glasses obtained with
cooling rates accessible in real experiments. For the inter-tetrahedral
angle Si-O-Si, the distribution shown in Fig.~9d has a maximum
around 130$^{\mathrm o}$, and a comparison to the corresponding
data for pure \silica\, glass shows that this position is shifted
to smaller angles due to the sodium atoms, in agreement with results
for a sodium silicate glasses~\cite{Ispas2001}.  Finally, we notice
that both $P_{\mathrm{SiONa}}$ and $P_{\mathrm{BONa}}$ distributions,
see Fig.~9g and h in Part~I, have maxima around 90$^{\mathrm o}$ and
present a significant decrease of the probability at large angles when
$T$ is lowered. These modifications are related to the fact that Na is
avoiding the direction of the Si-O bond (or B-O bond) as it can be seen
for example in Fig.~\ref{fig:fig6-b3na} for the trigonal units. The maxima
around 90$^{\mathrm o}$ are probably related to the angles corresponding
to   Na which share two oxygen atoms with the central cation (blue
circles in Figs.~\ref{fig:fig6-b3na}--\ref{fig:fig8-si4na})

Before we conclude this section on the structural properties of the glass,
we mention that we have not found any boroxol rings in our sample, and
this despite the relatively large concentration of \biii units. Thus,
in contrast to pure B$_2$O$_3$ glass in which these rings are relatively
frequent~\cite{Ferlat2008065504}, the presence of a second network
former Si and the network modifier Na make that these rings are no more
favorable structural units, in a compositional range with low content
of B$_2$O$_3$.

\section{\label{sec:vibpropr}Vibrational properties}

\begin{figure}
   \includegraphics[width=0.43\textwidth]{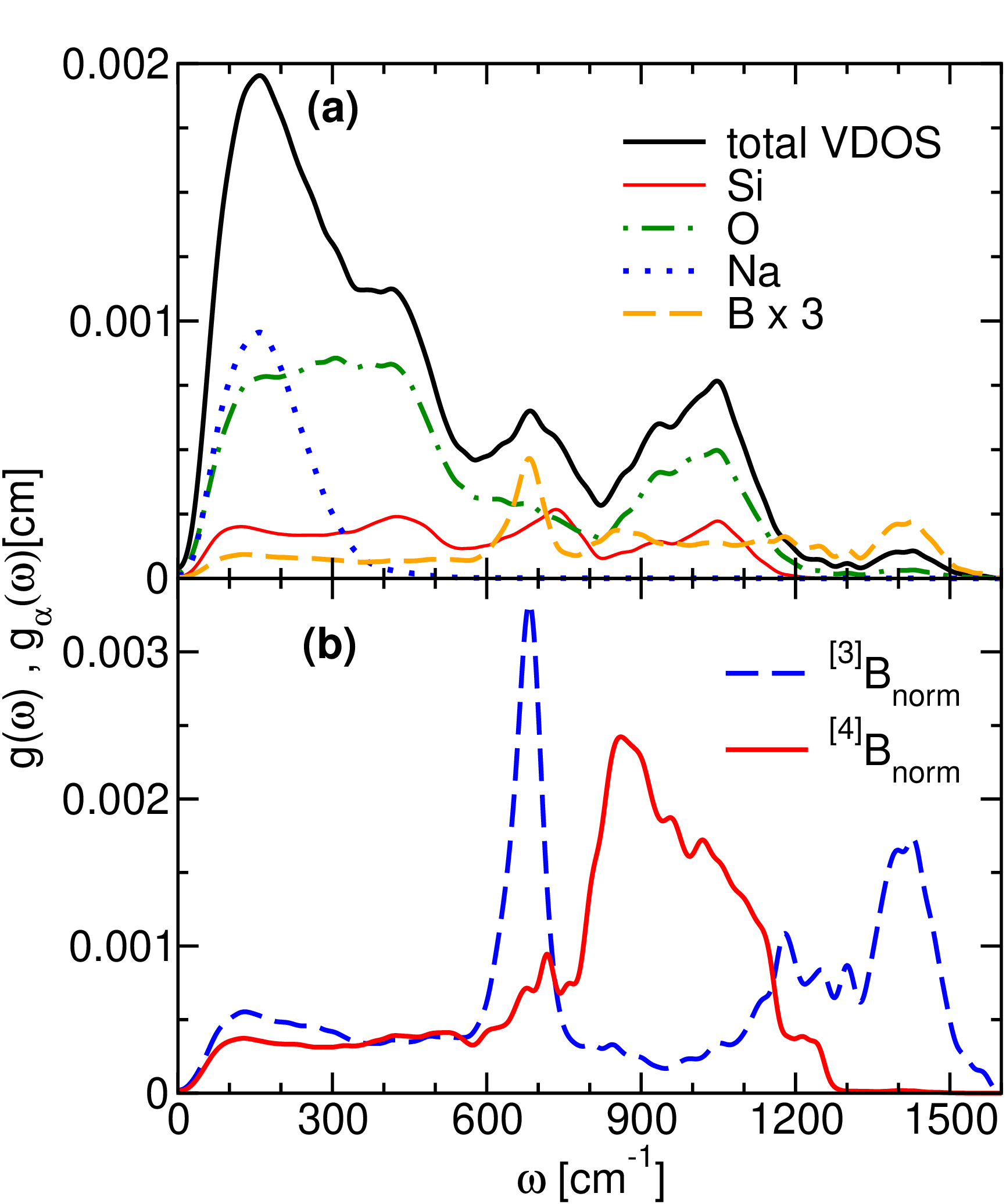}
    \caption{\label{fig:fig11-vdos1} (a) Total and partial vibrational
density of states for the glass. The total VDOS is normalized to 1,
and the partial VDOS of borons is multiplied by 3 in order to enhance
its visibility.  (b) Partial vibrational density of states for trigonal
$^{[3]}$B and tetragonal $^{[4]}$B boron atoms. Both are normalized
to 1 in order to allow a better comparison.}
\end{figure}

In order to study the vibrational properties of the NBS glass we
have relaxed its structure to 0~K and then determined the dynamical
matrix. This was done with a finite difference scheme using atomic
displacements of about~$2.6\cdot 10^{-3}\mathrm \AA$. The diagonalization
of the dynamical matrix gave then the $3N$ eigenvalues $\{\omega_p\}$
as well as the corresponding normalized eigenvectors ${\{\mathbf
e(\omega_{p})\}}$, for $p=1,\, 2,\ldots 3N$ where $N$ is the total number
of atoms in the sample, i.e.~320 in our case. The vibrational density
of states (VDOS) is then given by

\begin{equation}
\label{eq-VDOS}
g(\omega)=\frac{1}{3N-3}\sum_{p=4}^{3N}\delta(\omega-\omega_p) \quad ,
\end{equation}

\noindent
and it is shown in Fig. \ref{fig:fig11-vdos1}a. It is also instructive
to decompose the VDOS into the so-called partial-VDOS, i.e. the
contributions from the different species:

\begin{equation}
\label{eq-pVDOS}
g_{\alpha}(\omega)=\frac{1}{3N-3}\sum_{p=4}^{3N}\sum_{I=1}^{N_\alpha}\sum_{k=1}^3
|\mathbf e_{I,k}(\omega_p)|^{2} \delta(\omega-\omega_p) \quad ,
\end{equation}

\noindent
where $\alpha=$ Si, O, Na, and B and $\mathbf e_{I,k}(\omega_p)$
is that part of the eigenvector $\mathbf e(\omega_p)$ that contains
the 3 components of particle $I$. The so obtained discrete vibrational
spectra has been broadened with a Gaussian of FWHM of 30~cm$^{-1}$ and they
are included in Fig.~\ref{fig:fig11-vdos1}a as well.

The figure shows that in the total VDOS one can identify four main bands:
A first one which ranges from 0 to $\approx$~600~cm$^{-1}$, a second
one between 600~cm$^{-1}$ and 820~cm$^{-1}$, a third one between 820
and 1200 cm$^{-1}$, and finally the fourth one between 1200~cm$^{-1}$
and 1600~cm$^{-1}$. Inspection of the partial VDOS shows that the four
elements contribute in very different ways to these four bands. For the
band at the lowest frequencies we see that the Na atoms give rise to the
marked peak at around 180~cm$^{-1}$ and that the oxygen atoms contribute
significantly in the whole frequency range of this band, whereas the
contribution of the Si and B are relatively weak. These results are in
qualitative agreement with the ones obtained for sodium and lithium   
silicate glasses~\cite{Ispas2005,Du2006114702}. 

For frequencies below 1200~cm$^{-1}$, one finds that the
partial VDOS of Si and O present three main bands, as in pure
silica~\cite{Taraskin1997,Pasquarello1998,Giacomazzi2009064202}. However,
their features (position, width and intensity) are somewhat modified
due to the presence of Na and B. For example, in pure silica one
finds a prominent peak at around 800~cm$^{-1}$ which is related to Si
\cite{Pasquarello1998}, and our data shows that this peak is shifted
to lower frequencies and has an intensity similar to the one of the
band below 600~cm$^{-1}$, whereas in pure silica it is two times more
intense.  The so-called high-frequency band in pure silica located above
1000~cm$^{-1}$ \cite{Pasquarello1998} is also shifted to lower frequencies
and this  can be rationalized by recalling that the modes on the left
side of this band can be attributed to the softening of stretching motions
in silicates due to the depolymerization \cite{DeSousaMeneses201350}.

The presence of B in the system gives rise to a peak at around
680~cm$^{-1}$ and the band at the highest frequencies. (In the
latter one finds also contributions of the oxygen atoms that are
connected to B atoms.) These results agree with interpretations of
Raman and hyper-Raman experiments for alkali borates or borosilicate
glasses and melts \cite{Yano2003137,Simon2008155103,Manara2009777},
and are also supported by {\it ab initio} modelling of vibrational
spectra (VDOS, infra-red, Raman and hyper-Raman) for \boric\, glass
\cite{Umari2005137401,Ferlat2008065504}.  Due to the importance of
the $^{[3]}$B and $^{[4]}$B units, it is useful to decomposed the
partial VDOS of the boron atoms into contributions that come from
these two motifs and in Fig.~\ref{fig:fig11-vdos1}b we show the two
spectra. Note that in order to facilitate the comparison of the two
curves we have normalized them to one. Since we have found a strong
$T-$dependence of the concentration of the $^{[3]}$B and $^{[4]}$B units
this normalization also allows to infer how the real spectra would look
like if we would use a cooling rate that corresponds to the one in real
experiments, i.e. glasses with a concentration of $^{[3]}$B around 30\%
(see Subsec.~\ref{sec:PDFglass}). (Here we assume that the main effect
of the cooling rate is the concentration of the $^{[3]}$B and $^{[4]}$B
units and hence the other partial VDOS are affected only weakly.)

For frequencies below 600~cm$^{-1}$ the spectra for the  $^{[3]}$B
and $^{[4]}$B units are very similar and hence this part of the
spectrum should not be affected by cooling rate effects. At higher
frequencies we find that the vibration of the trigonal boron gives
rise to a pronounced peak at around 680~cm$^{-1}$ and a broad band
between 1100~cm$^{-1}$ and 1500~cm$^{-1}$.  This result might be
in contrast to the interpretation of Raman spectra in which one
relates the peak in the 600--700~cm$^{-1}$ region to the presence of
reedmergnerite and/or danburite rings  containing tetrahedral borons
\cite{Manara20092528,Manara2009777,Angeli2012054110}. Indeed we see
from our spectra that the $^{[4]}$B units do not have a significant
contribution at these frequencies (their vibrational modes are mainly
located in a band between 800~cm$^{-1}$ and 1200~cm$^{-1}$). Nevertheless
we cannot conclude that this experimental interpretation of the origin
of the peak  is incorrect, as in our decompositions we have considered
only the modes of the borons atoms in trigonal and tetrahedral units,
and not the modes of the whole reedmergnerite and/or danburite rings.
Finally we mention that the distinct features in the spectrum of the
\biii and \biv units allow to infer from the experimental spectrum the
concentration of these units in the sample.

\begin{figure}
   \includegraphics[width=0.43\textwidth]{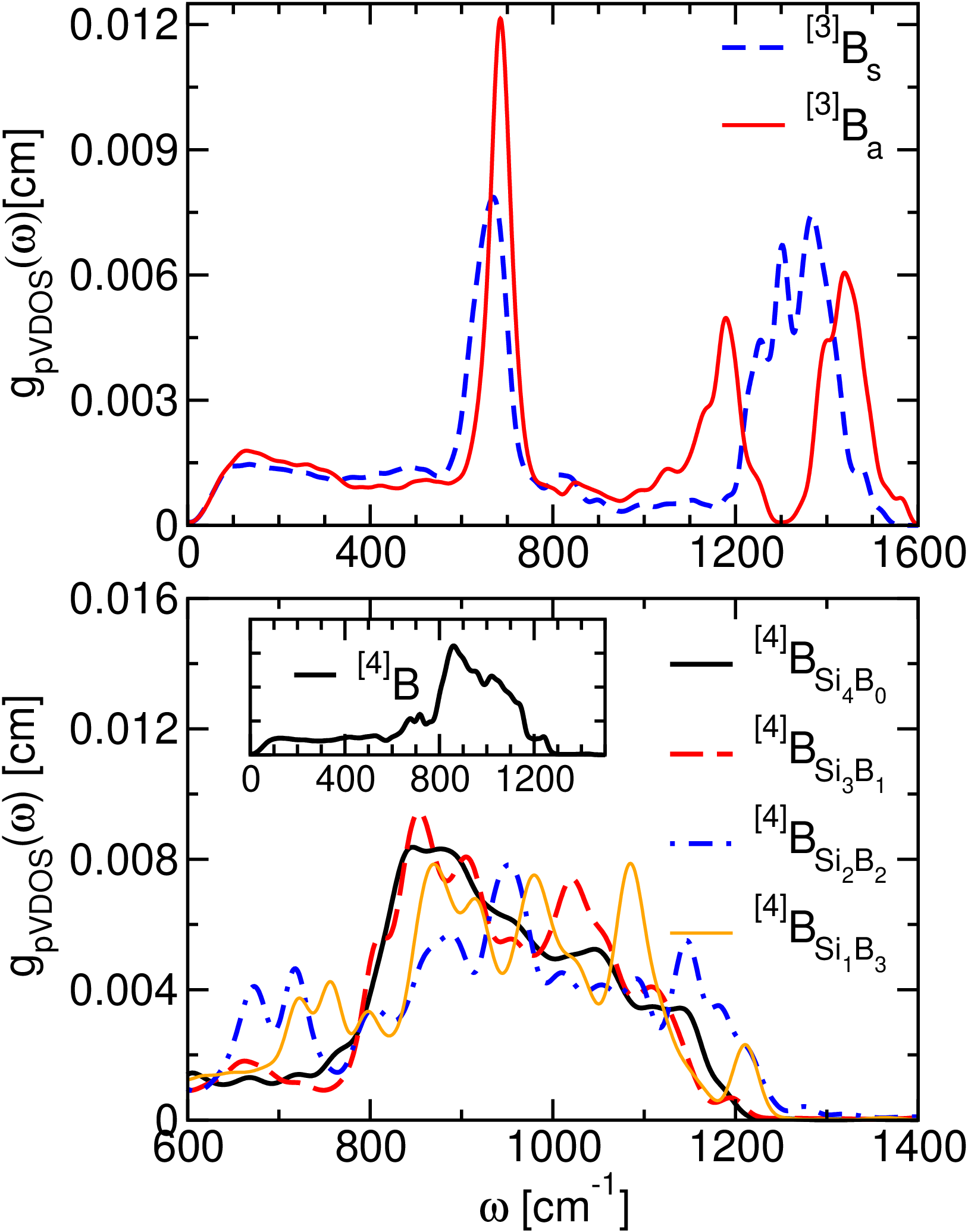}
    \caption{\label{fig:fig12-vdos2} Partial vibrational of state for boron.
(a) Decomposition of the pVDOS for \biii~into contributions from the
symmetric and asymmetric trigonal environments $^{[3]}$B$_s$ and
$^{[3]}$B$_a$; (b) Decomposition of the pVDOS for \biv~into contributions
related to different types of second nearest neighbors.}

\end{figure}

The vibrational features of the various atoms and species depend not
only on the number of their neighbors but also on how the latter are
connected to the rest of the network.  As discussed above, for the
trigonal boron atoms one can therefore distinguish between the symmetric
units ($^{[3]}$BO units completely connected to the network through BO
atoms), and the asymmetric units (boron atoms that have at least one NBO
as nearest neighbor). In Fig.~\ref{fig:fig12-vdos2}a we show the VDOS for
these two subunits and we recognize that the peak around 680 cm$^{-1}$,
has contributions from symmetric as well as asymmetric \biii units,
and that the peak of asymmetric units is more pronounced.

Also interesting is the high frequency band above 1100 cm$^{-1}$ since
the decomposition shows us that it is the sum of three peaks: Two peaks
located at around 1050 and 1450 cm$^{-1}$ which correspond to the modes of
the asymmetric \biii~and a broad peak located in between and which is related
to the vibrational motion of the symmetric \biii units.  Although we have
not investigate the details regarding what type of motion corresponds to
these three peaks it is likely that they can be to attributed to complex
stretching motions as it has been suggested for example from IR experimental
data for AgI-doped borate glasses \cite{Varsamis19993885}.

The decomposition of the \biv modes taking into account the nature of
the second network former to which the BO$_4$ unit is linked to is shown
in Fig.~\ref{fig:fig12-vdos2}b. (We recall that the large majority of
$^{[4]}$B have only BO as nearest neighbors.) Among the \biv units
present in our structures, the concentration of the reedmergnerite
and danburite suprastructural units  \cite{Manara20092528} (labelled
$^{[4]}$B$_{\rm Si_{\rm 4}B_{\rm 0}}$ and $^{[4]}$B$_{\rm Si_{3}B_{1}}$,
respectively) dominate and is around 43\% and 27\% respectively. We
see that these two structures give rise to a very similar partial
VDOS with a asymmetric band above 800~cm$^{-1}$.  This result may
contradict the interpretation of recent Raman investigations in which
the modes of the reedmergnerite and danburite suprastructural rings
were assigned to a different spectral band, between 600-700~cm$^{-1}$
\cite{Manara20092528,Manara2009777,Angeli2012054110}.  However,
as already stated above, one has to recall that our assignments are
based regarding the modes of tetrahedral borons, and not of the rings to
which their belong to, as done in experimental studies which make use of
analogies to the experimental spectra of reedmergnerite and danburite
crystals. For the structural units $^{[4]}$B$_{\rm Si_{2}B_{2}}$ and
$^{[4]}$B$_{\rm Si_{1}B_{3}}$ we find that the band above 800~cm$^{-1}$
is more symmetric but now we find also modes at around 700~cm$^{-1}$
and 750~cm$^{-1}$, respectively. Hence our analysis gives evidence that
this spectral band can give information on the nature of the second
nearest neighbor.  However, due to the rather poor statistics of the
present simulation it is hard to draw more quantitative statements.

\section{\label{sec:electronic} Electronic properties}

In this subsection we will analyze the electronic properties of our NBS
glass samples in terms of the electronic densities of states and the
Bader charge variations.

\subsection{\label{sec:edos}Electronic density of states }

To calculate $D(E)$, the average total electronic density of states
(eDOS), we have in a first step used the Kohn-Sham energies obtained for
each 0~K relaxed structure to  determine the partial eDOS $D_\alpha^l
(E)$ for $\alpha=$ Si, O, Na, B and the angular momenta $l=0,\,
1$. Subsequently $D(E)$ has been obtained as $D(E)=\sum_\alpha  c_\alpha
\sum_{l=0,1} D_\alpha^l (E)$, where $c_\alpha$ is the concentration of
species $\alpha$.

\begin{figure}
   \includegraphics[width=0.43\textwidth]{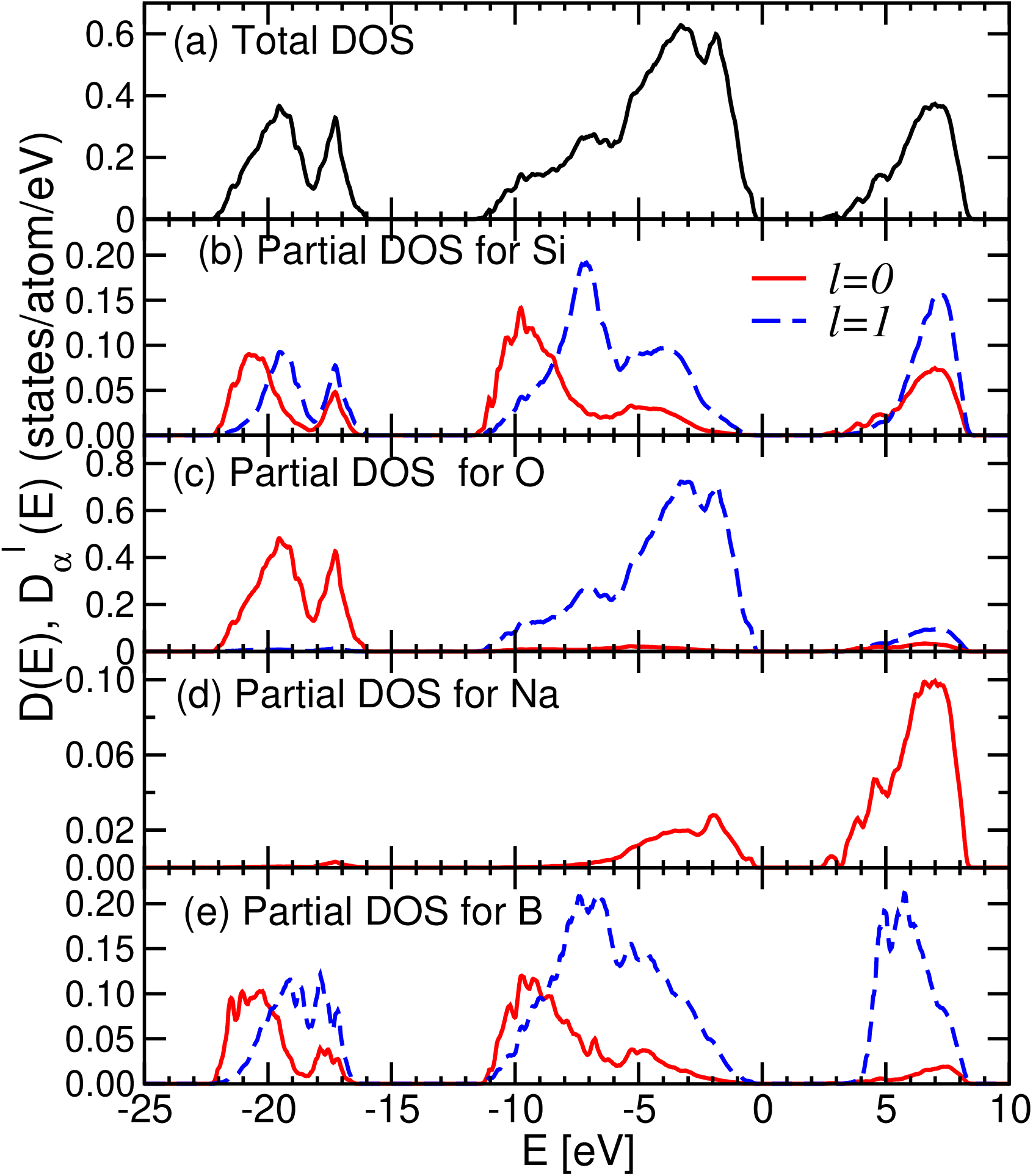}
    \caption{\label{fig:fig13-edos} (a) Total electronic density of states
$D(E)$ in the glass.  (b)-(e) Partial electronic density of states
$D_\alpha^l (E)$ for $\alpha=$Si, O, Na, and B.  The solid and dashed
lines show $D_\alpha^l (E)$  for $l=0$ and $l=1$, respectively. The
origin of the energy is the Fermi level.}
\end{figure}

In Fig.~\ref{fig:fig13-edos} we show $D(E)$ as well as the partial eDOS,
$D_\alpha^l (E)$, for $\alpha=$ Si, O, Na, B and $l=0,\, 1$ respectively.
Above the Fermi level, $E=0$, we find a band gap of approximately 3~eV.
Although to our knowledge, experiments have not yet determined the
value of this gap for NBS glasses, we can compare it to values for other
glasses. For the case of \silica, DFT calculations have predicted values
around 5~eV~\cite{Sarnthein199512690,Benoit2000,Du2006114702}, whereas
for sodium and lithium silicates the width of the gap was found smaller:
3.4~eV for the lithium disilicate~\cite{Du2006114702} and ranging
between 2.77 and 2.86~eV for sodium tetrasilicate~\cite{Ispas2001}.
This narrowing of the gap has been argued to be related to the presence
of the Na and Li atoms, and the same explanation is likely to hold also
for our NBS composition which presents a high soda content. However,
when making these arguments one should not forget the well known problem
of standard DFT calculations which typically predict gaps that are
substantially smaller than the experimental ones, both for semiconductors
and insulators \cite{Burke2012150901}.

Below the Fermi level, $D(E)$ shows two valence bands: A first one
between $-22$ and $-16$~eV and a second one between $-11$ and 0~eV.
The low energy band is also observed in the partial eDOS for Si, B, and O
and it has a double peak structure. Since the peak at higher energy,
$E \approx -17$~eV, is absent in fully connected networks such as pure
\silica\, and \boric, we can assign it to electronic states of Si and B
atoms which have at least one NBO, in agreement with results for lithium
silicate crystals and glasses~\cite{Du200622346,Du2006114702}, as well as for
a sodium silicate \cite{Ispas2001}. We also note that for Si and B this
double peak structure is present for $l=0$ and $l=1$, thus indicating
a hybridization of their respective $2s$ and $2p$ states with each other.

The features of the valence band at energies above $-11$~eV can be
understood by recalling the ones observed in pure \silica\, and \boric\,
glasses \cite{Sarnthein199512690,Ohmura2008224206}, in spite of the presence
of a small but negligible contribution in $D_{\mathrm{Na}}^l (E)$
(see Fig.~\ref{fig:fig13-edos}d).  Thus the band between $-11$~eV and
$-5$~eV may be assigned to contribution from bonding states between
O~$2p$ orbitals and  Si $sp^3$ hybrids on one hand, and with  hybridized
$2s$ and $2p$ orbitals of B atoms on the other hand. Finally there
is a large peak above $-5$~eV which originates mainly from O~$2p$
since in this energy range the contributions from $D_{\mathrm{Si}}^l
(E)$ and $D_{\mathrm{B}}^l (E)$ are substantially smaller. This
peak is generally assigned to lone-pair O non-bonding $2p$ states
\cite{Sarnthein199512690,Ohmura2008224206}.

\subsection{\label{sec:charge} Bader  charges}

Since simulations within a DFT approach allow to calculate the electronic
properties of the system, one can correlate these with the local
structure.  In the present study we have employed the method proposed by
Bader \cite{Bader_book}, which is also called ``atom in molecule''. In
this approach one partitions the charge density into regions around
each nucleus, the so-called Bader volume, using as boundary surfaces with 
zero flux, i.e. perpendicular to this surface the change density $\rho(
\mathbf r)$ is in a minimum: $\nabla \rho( \mathbf r)\cdot \mathbf n=0$
 \cite{Du200622346,Henkelman2006}. The charge is then given by

\begin{equation}
  \label{eq-bader}
  Q_{\alpha}^{Bader}=Z_{\alpha}-\int_{\rm Bader~volume}\rho(\mathbf r)dV   \quad ,
\end{equation}

\noindent
where $Z_\alpha$  is the number of   electrons of an atom of type $\alpha$.

\begin{table}
\centering
 \begin{tabular}{c|c|c|c|c}

Charge $[|e|]$ & Si &\Q 4 & \Q 3  & \Q 2 \\
 \hline
liquid & 3.097(0.057) &  3.122(0.046) &3.086(0.054) & 3.075(0.049)\\
glass &  3.146(0.021)  & 3.159(0.018) & 3.137(0.017) & 3.113(0.012) \\[1ex]

& O & BO  & NBO  & TBO \\
 \hline
liquid &  -1.560(0.047) &   -1.580(0.036) &  -1.510(0.033) & -1.575(0.059) \\
glass &  -1.584(0.031) & -1.597(0.021) & -1.540(0.014) &  -1.607(0.011)  \\
[1ex]

& B &  $^{[3]}$B  & $^{[4]}$B  & Na \\
 \hline
liquid &   2.285(0.050) &   2.277(0.069) &  2.303(0.052) & 0.822(0.024) \\
glass &  2.328(0.027)  &  2.308(0.020) &  2.346(0.019)&   0.830(0.027) \\
\hline 
 \end{tabular}
\caption{Average Bader charges for atoms and various species found in
NBS  liquid at 2200~K and in the glass at 0~K. In parenthesis we give the 
standard deviations.
}
\label{table-bader}
\end{table}

We have found that the so obtained charges are basically gaussian
distributed and hence they can be characterized by their mean value
and width.  The average Bader charge of silicon, oxygen, boron,
and sodium atoms as well as those of various species present in the
structure of the NBS liquid at 2200~K and in the glass at 0~K are listed
in Tab.~\ref{table-bader}. We recognize that these charges are quite
different from the nominal valence charges (i.e. +4 for Si, -2 for O,
+1 for Na and +3 for B) which indicates a mixed ionic-covalent nature
of the interactions. In the liquid, the absolute values of the charges
are smaller than in the glass state, and their standard deviations are
larger which shows that in the liquid state there is a higher degree
of disorder than in the glass state, in agreement with our analysis
for the structural quantities and results of previous \ab simulations
\cite{Du2006114702}. For Si we see that the charge of the \Q n species
increases with $n$, a trend that has been termed ``non-localized
effect'' of the sodium atoms~\cite{Bruckner198049,Uchino1991}
and which has also been found in other glassy and crystalline
silicates~\cite{Ispas2001,Du2006114702, Du200622346}.

For the borons we see that the charge depends on their coordination
number, with the $^{[4]}$B being more positive than the $^{[3]}$B. This
result is consistent with the observation  that the former structure is
surrounded by less Na atoms than the latter one (if one considers the
number of Na per BO).

Regarding the oxygen species, we find that the BO's are slightly more
negative that the NBOs, in both glass and liquid states. Note that
a formal charge neutrality requires a more negative charge of NBO's,
and previous DFT calculations on sodium and lithium silicates having
NBO's~\cite{Ispas2001, Du2006114702, Du200622346} have indeed shown this to
be the case. In fact this failure of the Bader method to describe
the different ionicity of Si-BO and Si-NBO bonds has already been
pointed out for lithium disilicate  \cite{Du2006114702,Du200622346}, as well
as in DFT calculations on clusters representing siliceous zeolites
\cite{Zwijnenburg2002}, and it originates from the partitioning scheme
of the total electron density.

\section{\label{sec:dielectric}{Dielectric properties}}

\subsection{\label{sec:Born-charge}{Born charge tensors}}
 Another possibility to associate an effective charge to a given atom
$I$ is to measure the force $\vec{F^I}$ that an external electric field
$\vec{\mathcal E}$ induces on the atom~\cite{Gonze199710335}. For this so-called
Born effective charge one thus measures the tensor

\begin{equation}
  \label{Z_ij}
Z_{I,ij}=\frac{\partial F^{I}_i}{e\partial \mathcal E_j}
  \quad  I=1,2, \ldots , N,   \quad  i, j \in\{x, y, z\} ,
\end{equation}

\noindent
where $e$ is the elementary charge. Since on average our the  system is
isotropic, and in fact we have found that the ensemble average values
of the off-diagonal elements of this tensor are negligible, one can
define the average Born charge $Z$ as a third of the trace of $Z_{I,ij}$
The so obtained charges are listed in Table~\ref{table-born-charge}.
If one compares these values with the corresponding Bader charges,
see Table~\ref{table-bader}, one see that most of them agree within the
standard deviation. The notable exception is Na, for which the Born charge
is significantly higher than the Bader charge (1.063$e$ vs. 0.822$e$
). This difference can be rationalized by the fact that the sodium atoms
are relatively mobile and hence will show a significant susceptibility,
i.e. high effective charge, when an external field is applied.

For the sake of comparison we have included in the table also the
values of the charges for pure silica glass (Si, O, BO) as obtained in
Refs.~\cite{Pasquarello19971766,Giacomazzi2009064202}.  We recognize
that for O and BO these charges are quite similar to the ones of the
NBS glass, thus showing that the presence of the additional elements
B and Na do not change these values in a significant manner. A larger
difference is found for the charge of the Si atoms which for the case
of NBS is smaller than the one in pure SiO$_2$. This decrease is likely
related to the fact that the local environment of Si has changed from an
almost perfect tetrahedral coordination to a distorted tetrahedral one,
due to presence of NBO atoms and Na neighborhood.

\begin{table}
\centering
 \begin{tabular}{c|c|c|c}
 \hline
    &   NBS & SiO$_2$ \cite{Giacomazzi2009064202} \quad &  \quad SiO$_2$ \cite{Pasquarello19971766}\quad \\
 \hline
Si  &  2.996(0.198) &  3.298(0.102)  &  3.177(0.121) \\
O   & -1.591(0.099) & -1.649(0.062)  & -1.588(0.078) \\
BO  & -1.634(0.074) & -1.649(0.062)  & -1.588(0.078) \\
NBO & -1.472(0.043) &  /             &  /             \\
Na  &  1.063(0.035) &  /             &  /             \\
B   &  2.142(0.245) &  /             &  /             \\
\biv & 2.398 (0.179) & /             &  /             \\
\biii &2.009 (0.153) & /             &  /             \\
\hline
\end{tabular}
\caption{
Average Born charge for the Si, O, Na, and B atoms. (In parenthesis are
the standard deviations). Also listed are the charges found in SiO$_2$
from Refs.~\protect\cite{Pasquarello19971766,Giacomazzi2009064202}.}
\label{table-born-charge}
\end{table}

\subsection{\label{sec:IRborn}Correlation between Born charges and structural features}

The knowledge of the structure and local charges allows us to investigate
the correlation between these two quantities. That such correlations do
exist has, e.g., been demonstrated for the case of silica for which it
has been found that the oxygen Born charge decreases as the corresponding
Si-O-Si angle increases~\cite{Pasquarello19971766,Giacomazzi2009064202}.
Since in NBS one has a much richer variety of local environments it is of
interest to see whether also for this system one can find correlations
between the charge  of an atoms and its local structure.

\begin{figure}
   \includegraphics[width=0.43\textwidth]{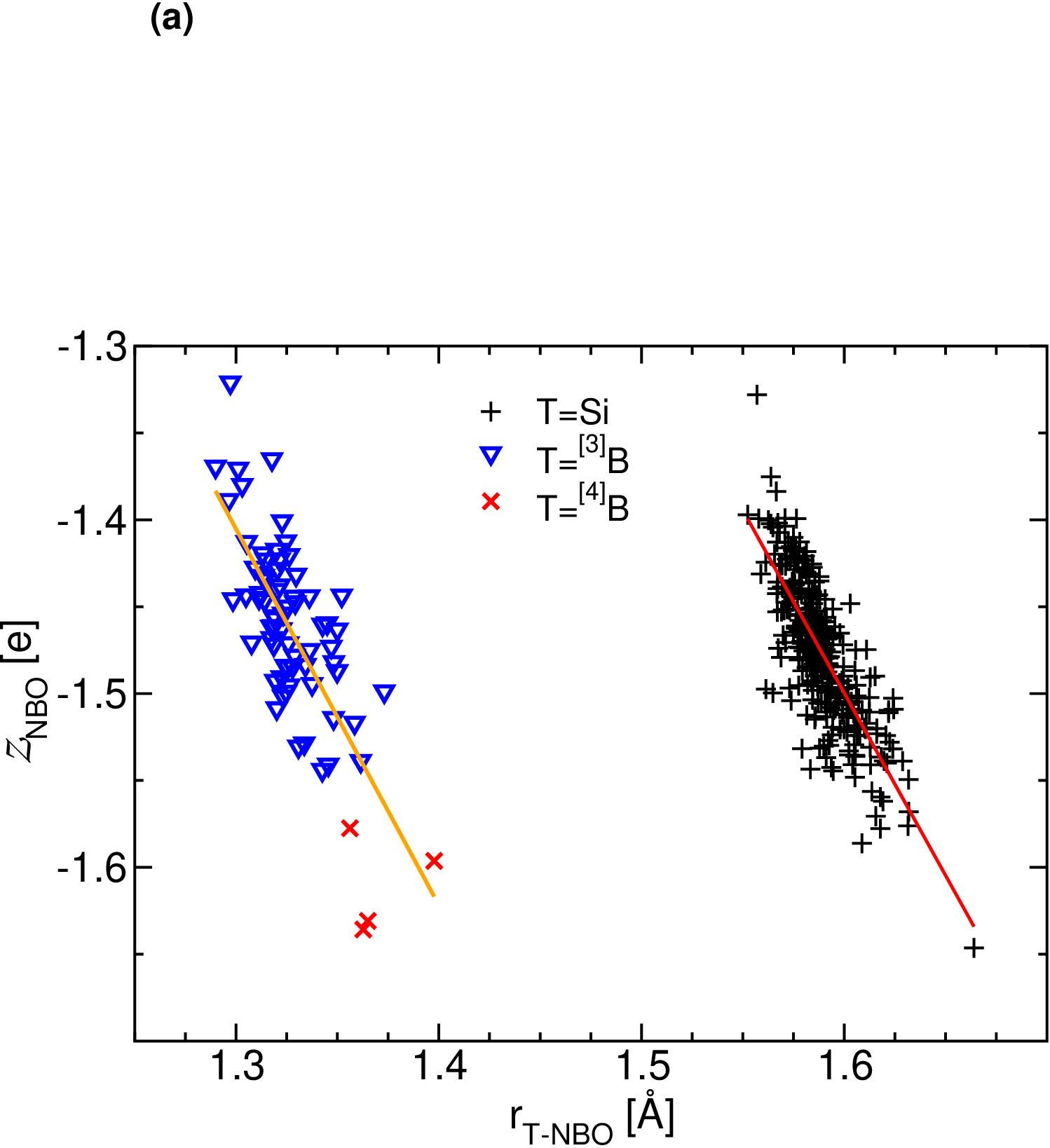}
    \caption{\label{fig:fig14-OBorn2bond}Dependence of the trace of the Born
charge tensor $Z^*_{\rm NBO}$ for a non-bridging oxygen atom on the bond length
with the network former $T$ for $T=$ Si, \biii and \biv. The straight
lines are the linear fits with the expressions given in Eqs.~(\ref{Z_SinbO})
and (\ref{Z_BnbO}).}
\end{figure}

First we investigate the relationship between the NBOs Born charge and
the associated bond length T-NBO, where T is the network former: T = Si,
$^{[3]}$B, and $^{[4]}$B, see Fig.~\ref{fig:fig14-OBorn2bond}.  We see
that there is indeed a significant correlation between these two quantities
in that the (absolute) value of the charge increases with increasing
distance, a result that is reasonable since with increasing distance
the charge on the atom becomes less screened. This correlation can be
fitted well with a linear function and hence we can write

 \begin{equation}
  \label{Z_SinbO}
   Z^{\rm Si}_{\rm NBO}[e]=-2.10e[\mathrm \AA^{-1}]
\times r_{\rm Si-NBO}[\mathrm \AA]+1.86\, e
\end{equation}

 \begin{equation}
  \label{Z_BnbO}
  Z^{\rm B}_{\rm NBO}[e]=-2.16e[\mathrm \AA^{-1}] 
\times r_{\rm B-NBO}[\mathrm \AA]+1.41\, e \quad .
 \end{equation}

\noindent Note that the very few NBO that are bounded to a \biv (red crosses)
fall more of less on the same line as the ones connected to a \biii
(blue triangles) which shows that this linear relation does not depend
strongly on the environment of the boron atom. Furthermore we see that
also the slope for the data for T = Si is very close to the one for
T = B from which we can conclude that the distance dependence of the
NBO charge is independent of the atom to which the oxygen is attached
(apart from a constant).

\begin{figure}
   \includegraphics[width=0.43\textwidth]{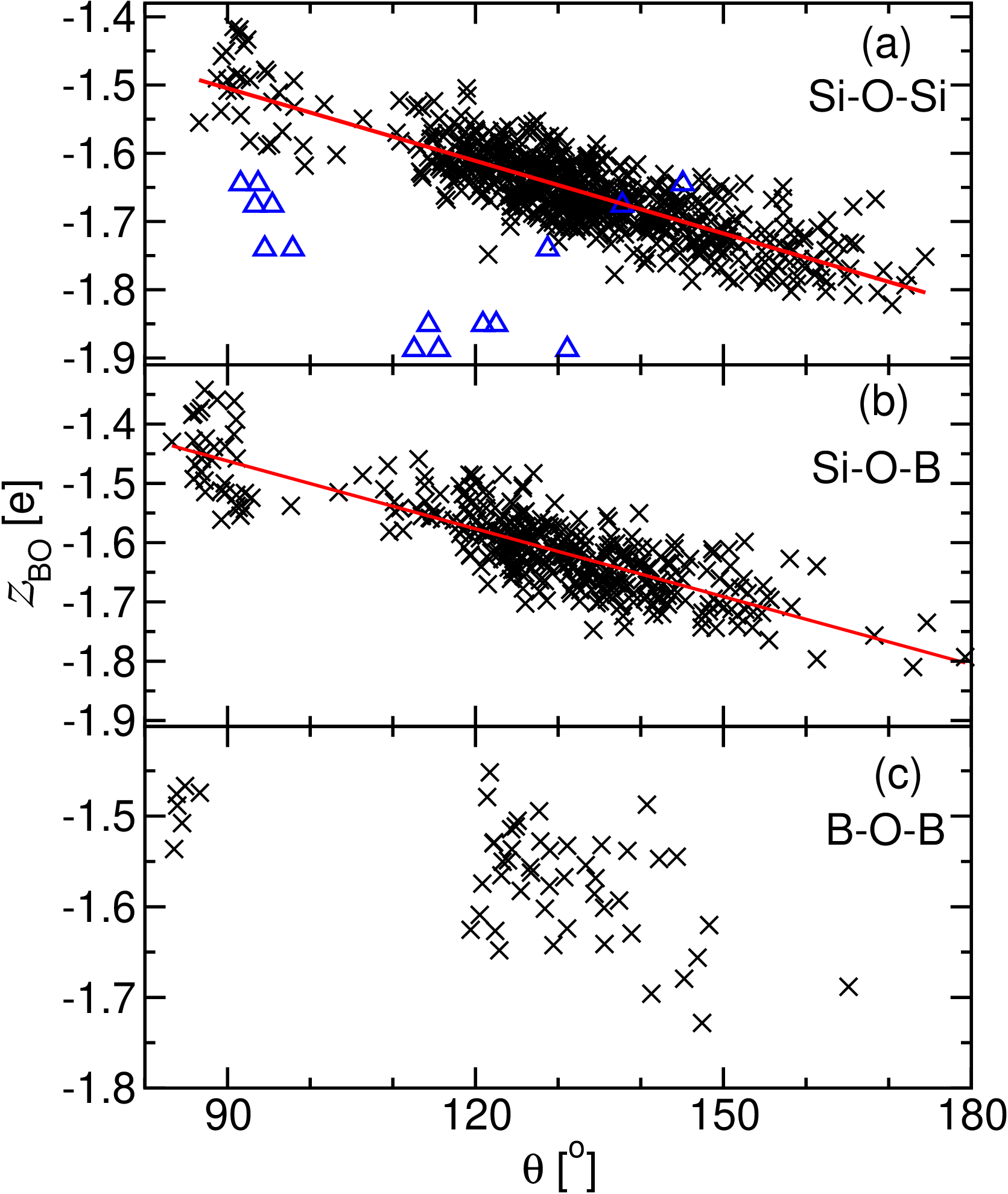}
    \caption{\label{fig:fig15-OBorn2angle}
Dependence of the trace of the
Born charge tensors $Z_{\rm BO}$ for a bridging oxygen atom on its bond angle
with the two network formers: (a) Si-O-Si, (b) Si-O-B, and (c) B-O-B. The
straight lines are the linear fits with the expressions given by Eqs. (\ref{Z_SiOSi}) and
(\ref{Z_SiOB}). The blue triangles in panel a) are the oxygen triclusters.}
\end{figure}

For bridging oxygens we have investigated the correlation between
$Z_{\rm BO}$ and the bond angle spanned by the two bonds to
its nearest neighbors (see \cite{Pasquarello19971766} for a similar
analysis in the case of silica). This correlation is shown in
Fig.~\ref{fig:fig15-OBorn2angle} (crosses) and we see that the charge
depends linearly on the angles Si-O-Si and Si-O-B. The linear fits shown
in the figure are given by

\begin{equation}
  \label{Z_SiOSi}
  Z^{\rm Si-O-Si}_{\rm BO} [e]=
     -0.0035e[\mathrm{deg}^{-1}]\times \theta_{\rm Si-O-Si}[\mathrm{deg}]-1.186\, e
\end{equation}

 \noindent  and 

\begin{equation}
  \label{Z_SiOB}
   Z^{\rm Si-O-B}_{\rm BO}[e] =
        - 0.0038e [\mathrm{deg}^{-1}] {\times} \theta_{\rm Si-O-B}[\mathrm{deg}]-1.119\, e \quad.
\end{equation}

\noindent
Thus we see that, within the accuracy of the data, the two slopes are the
same and only the mean value of the charge is slightly different. Also
included in panel a) are the charges/angles for oxygen triclusters (blue
triangles). We see that these charges are on average significantly more
negative than the ones for BO and that they do not follow the linear
relation given by Eq.~(\ref{Z_SiOSi}) and hence they have not been
considered in this fit. 

In all three panels we see some data points at angles around
90$^{o}$. These are related to defective structure that are present
because of the high quench rate and hence should not exist in the real
glass (see panels d) and e)  of Fig.~9  in Part I). Interestingly
enough, the linear fits give a good description also of these defective
structures, thus showing that the relation between angle and charge
is rather general. Finally we mention that for the B-O-B angle we do
not find a significant correlation between the angle and the charge,
at least if we do not take into account the defective structures with
angles around 90$^{o}$. (Note that the observed range in the value
of this angle is smaller than the one for Si-O-Si or Si-O-B. This is
probably not a real feature but is likely related to the fact that we
have only very few of these angles.)

 \subsection{\label{sec:dielconst}Dielectric constants}

The dielectric properties of our eight NBS glass samples have been
computed by using the density functional perturbation theory features
of the VASP package within the PAW methodology~\cite{Gajdos2006045112}.
For this we have first determined the purely electronic dielectric
tensor $(\epsilon_\infty)_{ij}$ (also called relative permittivity). As
expected for an amorphous material, we find that this tensor is basically
isotropic and diagonal and hence we define an orientationally averaged
dielectric constant $\epsilon_\infty$ (also called high frequency
dielectric constant) as one third of the trace of this dielectric
tensor. The so obtained value for $\epsilon_\infty$ is 2.4. This is
somewhat higher than the value obtained by Pasquarello and coworkers
for the case of pure silica who, using also DFT, found 2.0 and
2.1~\cite{Pasquarello19971766,Giacomazzi2009064202}, values that agree
well with experimental result equal to 2.1~\cite{Philipp1998}. Although
to our knowledge for NBS glasses with the composition considered here
the experimental value of $\epsilon_\infty$ has not yet been measured,
we can exploit the fact that the high-frequency dielectric constant
is related to the refractive index $n_{\infty}$ of the material,
$\epsilon_\infty=n^2_\infty$, which leads for our NBS glass to a value
of $n_{\infty} \approx 1.55$. This value can be compared with the
experimental result of 1.52 for the far-infrared refractive index of a
sodium borosilicate glass with a composition quite similar to our glass,
i.e. 27.8\% Na$_2$O--11.1\% B$_2$O$_3$--61.1 SiO$_2$, corresponding to
$R=2.5$ and $K=5.5$ \cite{Kamitsos199431}. From the good agreement between
these two values for $n_{\infty}$ we can conclude that this quantity
does probably not depend strongly on the composition of the glass, in
agreement with experimental findings for series of sodium borosilicates
and aluminosilicates glasses \cite{Kamitsos199431,Hsieh19961704}.

Furthermore it is also possible to calculate the static dielectric
constant $\epsilon_0$ which reflects the ionic displacement
contributions to the dielectric constant. For this one makes use of
the Born effective charges introduced above as well as the vibrational
modes and eigenfrequencies, the latter ones have been obtained as
explained in Sec.~\ref{sec:vibpropr}. $\epsilon_0$ can then be expressed
as \cite{Pasquarello19971766}:

\begin{equation}
  \label{epsilon0}
\epsilon_0 =\epsilon_\infty + 
            \frac{4\pi}{3V} \sum_p \sum_j \frac{|\mathcal F_j^p|^2}{\omega_p} \quad ,
\end{equation}

\noindent
where the so-called oscillator strength $\mathcal F^{p}_j$ is  defined as

\begin{equation}
\mathcal F_{j}^{p} = 
\sum_{I,k} 
Z_{I,jk} \frac{{\mathbf e_{I,k} (\omega_p)}}{\sqrt{m_{I}}}  \, .
\end{equation}

Using our value for the high-frequency dielectric constant,
$\epsilon_\infty=2.40$, the above formula gives an average dielectric
constant $\epsilon_0$ of 6.62.
This result compares well with  experimental values for  more complex sodium borosilicate glasses with low sodium content  for which $\epsilon_0$ ranges between 5.5
and 6.1 \cite{Wang20081128}.
Furthermore we mention that for the
case of pure silica the DFT calculations predict values between 3.6
and 3.8~\cite{Pasquarello19971766,Giacomazzi2009064202}, results that
are close to the experimental findings. Hence we can conclude that,
in contrast to the high frequency dielectric constant for which we
have found only a relatively weak dependence on the glass composition,
$\epsilon_0$ depends quite strongly on the type of glass considered.

\subsection{\label{sec:IRdielec}Dielectric functions: infrared spectra }

A further observable of interest is the frequency dependence
of the dielectric function $\epsilon(\omega)$. This complex
quantity can be calculated directly from the eigenmodes and Born
effective charges: Writing $\epsilon(\omega)=\epsilon_1(\omega)+
i\epsilon_2(\omega)$, the real and imaginary parts are given by
\cite{Thorpe19868490,Pasquarello19971766}:

\begin{eqnarray}
\epsilon_{1}(\omega) &= &
\epsilon_{\infty}-\frac{4\pi}{3V}\sum_{p} \sum_j 
\frac{\mid \mathcal F_j^{p}\mid^2}{\omega^2-\omega_{p}^2}  \\
\epsilon_{2}(\omega) &= &
\frac{4\pi^2}{3V}\sum_{p} \sum_j
\frac{\mid \mathcal F_j^{p}\mid^2}{2\omega_{p}^2} \delta(\omega-\omega_{n}). 
\end{eqnarray} 

Closely related to $\epsilon(\omega)$ is the absorption spectra
$\alpha(\omega)$ which is given by~\cite{Kamitsos199431}

\begin{equation}
\alpha(\omega) = 4\pi\omega n''(\omega)  \quad ,
\label{eq_absorption_1}
\end{equation}

\noindent
with
\begin{equation}
n''(\omega) = \sqrt{\frac{\sqrt{\epsilon_1^2+ \epsilon_2^2}- \epsilon_1}{2}}\quad .
\label{eq_absorption_2}
\end{equation}

\noindent
This function can be measured directly in experiments and hence will
allow us to make a comparison with the prediction from the simulation.

\begin{figure}
   \includegraphics[width=0.43\textwidth]{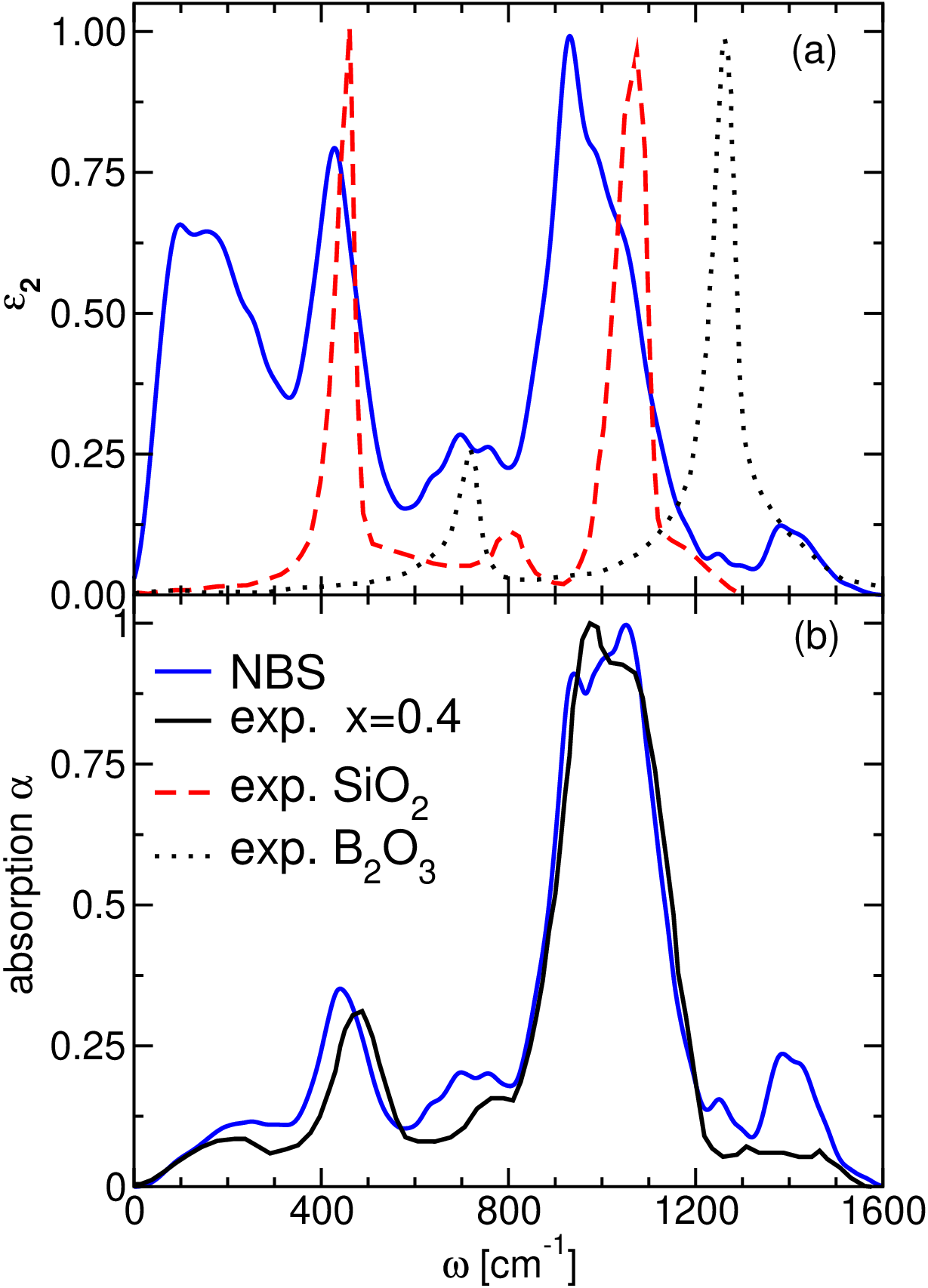}
    \caption{\label{fig:fig16-IRdielec} Imaginary part of the dielectric
function $\epsilon_2 (\omega)$ and  absorption spectrum $\alpha(\omega)$, panel (a) and
(b) respectively. Also included are the experimental spectra for pure
\silica\, and \boric\,     glasses (panel a) \protect\cite{GaleenerFL,Philipp1998}.
For the absorption spectrum $\alpha(\omega)$ (panel b), we have also included the experimental spectra
for a glass of composition Na$_2$O-$x$ B$_2$O$_3$-($3-2x$) SiO$_2$ for
$x=0.4$ from Ref.~\protect\cite{Kamitsos199431}. For the sake of comparison we
have normalized the maximum amplitude of each spectrum to unity.}
\end{figure}

In Fig. \ref{fig:fig16-IRdielec} we show the imaginary part of the
dielectric function $\epsilon_2  (\omega)$ and the absorption spectra
calculated for our NBS glass samples (panels a) and b) respectively).
In panel a) we have also included the experimental spectra for
\silica\,  and \boric~glasses~\cite{GaleenerFL,Philipp1998}, and in
panel b) the experimental absorption spectrum obtained for a sodium
borosilicate glass with ratios $R=2.5$ and $K=5.5$, i.e. of composition
Na$_2$O-$x$B$_2$O$_3$-($3-2x$) SiO$_2$ with $x=0.4$~\cite{Kamitsos199431}.
For the sake of comparison we have scaled the maximum amplitude of each
curve to 1.0.

The comparison of the spectra for the three glass formers shown in
Fig.~\ref{fig:fig16-IRdielec}a allows to identify the origin of the
various peaks. Firstly, the spectrum for NBS shows a broad band below
300 cm${^{-1}}$. Since this band is absent in the two other spectra and
since we have seen in the VDOS (see Fig.~\ref{fig:fig11-vdos1}) that in
this frequency range the vibrational motion is dominated by the Na atoms
we can conclude that the band is related to the sodium atoms.

Furthermore we notice the presence of a narrow band slightly above 400
cm${^{-1}}$. This feature is also present in the spectrum of pure silica
and is known to originate from the bending and rocking motion of oxygen
atoms~\cite{Taraskin1997,Pasquarello1998}. The fact that for NBS the
peak is broadened and also shifted somewhat to lower frequencies is
probably related to the slight opening of the Si-O-Si bond angles and
a softening of the effective Si-O interactions due to the presence of
the network modifiers.

The next band is found between 600 and 800 cm${^{-1}}$. In this
frequency range the experiments on \boric\,  and SiO$_2$ glasses
show a peak, the origin of which is attributed to the bridging
oxygen~\cite{Kamitsos199431,Manara2009777}. Although we find that in
NBS the corresponding band is at somewhat lower frequencies, it can be
expected that it has the same origin and that the red shift is due to
the presence of the network modifier.

At even higher frequencies we can distinguish two bands:
The first one ranges from 850 to 1200 cm${^{-1}}$, and it
can be assigned to oxygen stretching modes of Si-NBO and Si-BO
\cite{FDomine01,CMerzbacher01}. Earlier analysis of the partial VDOS
of a sodium tetrasilicate glass have shown that the main effect of
the presence of NBO atoms and the resulting depolymerization of the
silica network is the shift of this band to lower frequencies in the
NBS spectrum~\cite{Ispas2005}. The second high frequency band extents
between 1200 cm${^{-1}}$ and 1600 cm${^{-1}}$ and, as discussed in
the context of Fig. \ref{fig:fig12-vdos2}, it arises from motions of
oxygen atoms belonging to borate units, namely the symmetric trigonal
ones $^{[3]}$B. This interpretation is also supported by the fact
that the experimental data for B$_2$O$_3$, which has mainly $^{[3]}$B
units, does have a pronounced peak at these high frequencies as well
(included in Fig.~\ref{fig:fig16-IRdielec}b as well).

Finally we compare our calculated absorption spectra with the experimental
data as measured by Kamitsos et al. \cite{Kamitsos199431} for the
borosilicate described above. From Fig.~\ref{fig:fig16-IRdielec}b we
can conclude that in general there is a rather good agreement between
the results from the simulations and the experimental data:  At low and
intermediate frequencies the position and height of the bands are 
well reproduced. The observed discrepancies can be rationalized by recalling
that our samples have a lack of of tetrahedral borate units (see the discussion in the context of
Fig.~\ref{fig:fig2-b3b4-to-Tg})  which are expected to contribute to the
IR signal in this frequency range~\cite{Kamitsos199759}, in agreement
with our findings from Fig.~\ref{fig:fig11-vdos1} where we showed that in
this frequency range the contribution of $^{[4]}$B is much more important
than the one of $^{[3]}$B. 
At higher frequencies there is a very good agreement for the band between 800 and 1200~cm${^{-1}}$ inferring that our models reproduce accurately the streching motions \cite{DeSousaMeneses201350} in silicates.

Finally we see that also the intensity of the band between 1200 and
1600~cm${^{-1}}$ is higher than the one of the experiments, but that its
position matches the one seen in the real data. Also this discrepancy can
likely be traced back to the too high concentration of $^{[3]}$B units,
since, see Fig.~\ref{fig:fig11-vdos1}, they are the only ones that give
rise to the density of states in that frequency range.

\section{\label{sec:conclu}Summary and Conclusions}

Using {\it ab initio} molecular dynamics computer simulations we have
investigated the structural, electronic, and vibrational properties of the
ternary sodium borosilicate glass 3Na$_2$O-B$_2$O$_3$-6SiO$_2$. Particular
attention has been given to the coordination of the boron atoms which
are found to be present in trigonal and tetrahedral geometries, as
expected for this composition. We find that the concentration of these
two structural units depends significantly on the fictive temperature of
the glass (with \biii decreasing with decreasing $T_{\rm fic}$), thus affecting
the vibrational density of states considerably.  However, if one takes
into account the measured $T-$dependence of these units, one finds
concentrations that are in quantitative agreement with the model proposed
by Yun, Dell and Bray and with estimates from spectroscopic measurements.

Despite the high cooling rate with which the glass sample has been
produced, the accuracy of the interactions allows us to study the
local structure of the various structural units. In particular we have
investigated how the Na atoms are distributed around the \biii triangles
and \biv tetrahedra. From this analysis we can rationalize why the first
peak of the radial distribution functions between B and Na is broadened
and also understand why the number of Na that share one/two oxygen atoms
with the central B atom depends on whether one considers \biii or \biv
structures. Furthermore we see from these Na distributions that the nature
of a \biv tetrahedron is different from a SiO$_4$ tetrahedron in that the
former gives rise to a distribution that is significantly more structured.

From the partial vibrational density of states we can conclude that the
\biii and \biv units have very different signatures. Furthermore we have
shown that these spectra also allow to distinguish clearly whether a \biii
unit is symmetric or asymmetric. For the \biv we show that the spectrum
can also be used to gain information on the nature of the second nearest
neighbor (i.e. whether it is a B or Si atom).

We have also calculate the dielectric function $\epsilon(\omega)$ and
the absorption spectra. The latter is in good quantitative agreement
with experimental data, thus showing that the simulation is reliable
also for this type of observable.

Finally we have investigated the correlation between the Born effective  charge of an
oxygen atom and its surrounding  geometry. We have found that for the
non-bridging oxygen, there is a linear correlation of this charge with
the distance from the associated Si/B atom. For the bridging oxygen we
find a linear dependence of the charge with the angle spanned by its
two bonds. These charge dependencies show that for this type of glasses
it is unlikely that a rigid ion model will be able to give a reliable
description of the local structure since such models do not take into
account the observed charge transfer.

\acknowledgements

We thank D. R. Neuville and B. Hehlen for  stimulating discussions
on this work. Financial support from the Agence Nationale de la Recherche
under project POSTRE is acknowledged. 
This work was performed using HPC resources from GENCI (TGCC/CINES/IDRIS) (Grants
x2010095045, x2011095045 and x2012095045), and also on the HPC@LR cluster, Montpellier, France. 
 W. Kob acknowledges
support from the Institut Universitaire de France.

\clearpage


\begin{thebibliography}{10}%
\makeatletter
\providecommand \@ifxundefined [1]{%
 \ifx #1\undefined \expandafter \@firstoftwo
 \else \expandafter \@secondoftwo
\fi
}%
\providecommand \@ifnum [1]{%
 \ifnum #1\expandafter \@firstoftwo
 \else \expandafter \@secondoftwo
\fi
}%
\providecommand \enquote [1]{``#1''}%
\providecommand \bibnamefont  [1]{#1}%
\providecommand \bibfnamefont [1]{#1}%
\providecommand \citenamefont [1]{#1}%
\providecommand\href[0]{\@sanitize\@href}%
\providecommand\@href[1]{\endgroup\@@startlink{#1}\endgroup\@@href}%
\providecommand\@@href[1]{#1\@@endlink}%
\providecommand \@sanitize [0]{\begingroup\catcode`\&12\catcode`\#12\relax}%
\@ifxundefined \pdfoutput {\@firstoftwo}{%
 \@ifnum{\z@=\pdfoutput}{\@firstoftwo}{\@secondoftwo}%
}{%
 \providecommand\@@startlink[1]{\leavevmode}%
 \providecommand\@@endlink[0]{}%
}{%
 \providecommand\@@startlink[1]{%
  \leavevmode
  \pdfstartlink
   attr{/Border[0 0 1 ]/H/I/C[0 1 1]}%
   user{/Subtype/Link/A<</Type/Action/S/URI/URI(#1)>>}%
  \relax
 }%
 \providecommand\@@endlink[0]{\pdfendlink}%
}%
\providecommand \url  [0]{\begingroup\@sanitize \@url }%
\providecommand \@url [1]{\endgroup\@href {#1}{\urlprefix}}%
\providecommand \urlprefix [0]{URL }%
\providecommand \Eprint[0]{\href }%
\@ifxundefined \urlstyle {%
  \providecommand \doi [1]{doi:\discretionary{}{}{}#1}%
}{%
  \providecommand \doi [0]{doi:\discretionary{}{}{}\begingroup
  \urlstyle{rm}\Url }%
}%
\providecommand \doibase [0]{http://dx.doi.org/}%
\providecommand \Doi[1]{\href{\doibase#1}}%
\providecommand \bibAnnote [3]{%
  \BibitemShut{#1}%
  \begin{quotation}\noindent
    \textsc{Key:}\ #2\\\textsc{Annotation:}\ #3%
  \end{quotation}%
}%
\providecommand \bibAnnoteFile [2]{%
  \IfFileExists{#2}{\bibAnnote {#1} {#2} {\input{#2}}}{}%
}%
\providecommand \typeout [0]{\immediate \write \m@ne }%
\providecommand \selectlanguage [0]{\@gobble}%
\providecommand \bibinfo [0]{\@secondoftwo}%
\providecommand \bibfield [0]{\@secondoftwo}%
\providecommand \translation [1]{[#1]}%
\providecommand \BibitemOpen[0]{}%
\providecommand \bibitemStop [0]{}%
\providecommand \bibitemNoStop [0]{.\EOS\space}%
\providecommand \EOS [0]{\spacefactor3000\relax}%
\providecommand \BibitemShut [1]{\csname bibitem#1\endcsname}%
\bibitem{Varshneya_book}%
  \BibitemOpen
  \bibfield{author}{%
  \bibinfo {author} {\bibfnamefont{A.}~\bibnamefont{Varshneya}},\ }%
  \emph{\bibinfo {title} {Fundamentals of inorganic glasses, 2nd edition}}\
  (\bibinfo {publisher} {Society of Glass Technology},\ \bibinfo {year}
  {2006})%
  \bibAnnoteFile{NoStop}{Varshneya_book}%
\bibitem{Yun01}%
  \BibitemOpen
  \bibfield{author}{%
  \bibinfo {author} {\bibfnamefont{Y.}~\bibnamefont{Yun}}\ and\ \bibinfo
  {author} {\bibfnamefont{P.}~\bibnamefont{Bray}},\ }%
  \bibfield{journal}{%
  \bibinfo {journal} {J. Non-Cryst. Solids}\ }%
  \textbf{\bibinfo {volume} {27}},\ \bibinfo {pages} {363 } (\bibinfo {year}
  {1978})%
  \bibAnnoteFile{NoStop}{Yun01}%
\bibitem{Yun02}%
  \BibitemOpen
  \bibfield{author}{%
  \bibinfo {author} {\bibfnamefont{Y.}~\bibnamefont{Yun}}, \bibinfo {author}
  {\bibfnamefont{S.}~\bibnamefont{Feller}},\ and\ \bibinfo {author}
  {\bibfnamefont{P.}~\bibnamefont{Bray}},\ }%
  \bibfield{journal}{%
  \bibinfo {journal} {J. Non-Cryst. Solids}\ }%
  \textbf{\bibinfo {volume} {33}},\ \bibinfo {pages} {273} (\bibinfo {year}
  {1979})%
  \bibAnnoteFile{NoStop}{Yun02}%
\bibitem{Dell19831}%
  \BibitemOpen
  \bibfield{author}{%
  \bibinfo {author} {\bibfnamefont{W.}~\bibnamefont{Dell}}, \bibinfo {author}
  {\bibfnamefont{P.}~\bibnamefont{Bray}},\ and\ \bibinfo {author}
  {\bibfnamefont{S.}~\bibnamefont{Xiao}},\ }%
  \bibfield{journal}{%
  \bibinfo {journal} {J. Non-Cryst. Solids}\ }%
  \textbf{\bibinfo {volume} {58}},\ \bibinfo {pages} {1} (\bibinfo {year}
  {1983})%
  \bibAnnoteFile{NoStop}{Dell19831}%
\bibitem{Parkinson20074076}%
  \BibitemOpen
  \bibfield{author}{%
  \bibinfo {author} {\bibfnamefont{B.}~\bibnamefont{Parkinson}}, \bibinfo
  {author} {\bibfnamefont{D.}~\bibnamefont{Holland}}, \bibinfo {author}
  {\bibfnamefont{M.}~\bibnamefont{Smith}}, \bibinfo {author}
  {\bibfnamefont{A.}~\bibnamefont{Howes}},\ and\ \bibinfo {author}
  {\bibfnamefont{C.}~\bibnamefont{Scales}},\ }%
  \bibfield{journal}{%
  \bibinfo {journal} {J. Non-Cryst. Solids}\ }%
  \textbf{\bibinfo {volume} {353}},\ \bibinfo {pages} {4076 } (\bibinfo {year}
  {2007})%
  \bibAnnoteFile{NoStop}{Parkinson20074076}%
\bibitem{Parkinson2007415114}%
  \BibitemOpen
  \bibfield{author}{%
  \bibinfo {author} {\bibfnamefont{B.~G.}\ \bibnamefont{Parkinson}}, \bibinfo
  {author} {\bibfnamefont{D.}~\bibnamefont{Holland}}, \bibinfo {author}
  {\bibfnamefont{M.~E.}\ \bibnamefont{Smith}}, \bibinfo {author}
  {\bibfnamefont{A.~P.}\ \bibnamefont{Howes}},\ and\ \bibinfo {author}
  {\bibfnamefont{C.~R.}\ \bibnamefont{Scales}},\ }%
  \bibfield{journal}{%
  \bibinfo {journal} {J. Phys.: Condens. Matter}\ }%
  \textbf{\bibinfo {volume} {19}},\ \bibinfo {pages} {415114} (\bibinfo {year}
  {2007})%
  \bibAnnoteFile{NoStop}{Parkinson2007415114}%
\bibitem{Parkinson20081936}%
  \BibitemOpen
  \bibfield{author}{%
  \bibinfo {author} {\bibfnamefont{B.}~\bibnamefont{Parkinson}}, \bibinfo
  {author} {\bibfnamefont{D.}~\bibnamefont{Holland}}, \bibinfo {author}
  {\bibfnamefont{M.}~\bibnamefont{Smith}}, \bibinfo {author}
  {\bibfnamefont{C.}~\bibnamefont{Larson}}, \bibinfo {author}
  {\bibfnamefont{J.}~\bibnamefont{Doerr}}, \bibinfo {author}
  {\bibfnamefont{M.}~\bibnamefont{Affatigato}}, \bibinfo {author}
  {\bibfnamefont{S.}~\bibnamefont{Feller}}, \bibinfo {author}
  {\bibfnamefont{A.}~\bibnamefont{Howes}},\ and\ \bibinfo {author}
  {\bibfnamefont{C.}~\bibnamefont{Scales}},\ }%
  \bibfield{journal}{%
  \bibinfo {journal} {J. Non-Cryst. Solids}\ }%
  \textbf{\bibinfo {volume} {354}},\ \bibinfo {pages} {1936 } (\bibinfo {year}
  {2008})%
  \bibAnnoteFile{NoStop}{Parkinson20081936}%
\bibitem{Manara20092528}%
  \BibitemOpen
  \bibfield{author}{%
  \bibinfo {author} {\bibfnamefont{D.}~\bibnamefont{Manara}}, \bibinfo {author}
  {\bibfnamefont{A.}~\bibnamefont{Grandjean}},\ and\ \bibinfo {author}
  {\bibfnamefont{D.~R.}\ \bibnamefont{Neuville}},\ }%
  \bibfield{journal}{%
  \bibinfo {journal} {J. Non-Cryst. Solids}\ }%
  \textbf{\bibinfo {volume} {355}},\ \bibinfo {pages} {2528 } (\bibinfo {year}
  {2009})%
  \bibAnnoteFile{NoStop}{Manara20092528}%
\bibitem{Manara2009777}%
  \BibitemOpen
  \bibfield{author}{%
  \bibinfo {author} {\bibfnamefont{D.}~\bibnamefont{Manara}}, \bibinfo {author}
  {\bibfnamefont{A.}~\bibnamefont{Grandjean}},\ and\ \bibinfo {author}
  {\bibfnamefont{D.~R.}\ \bibnamefont{Neuville}},\ }%
  \bibfield{journal}{%
  \bibinfo {journal} {Amer. Mineral.}\ }%
  \textbf{\bibinfo {volume} {94}},\ \bibinfo {pages} {777 } (\bibinfo {year}
  {2009})%
  \bibAnnoteFile{NoStop}{Manara2009777}%
\bibitem{Pedesseau-nbs1}%
  \BibitemOpen
  \bibfield{author}{%
  \bibinfo {author} {\bibfnamefont{L.}~\bibnamefont{Pedesseau}}, \bibinfo
  {author} {\bibfnamefont{S.}~\bibnamefont{Ispas}},\ and\ \bibinfo {author}
  {\bibfnamefont{W.}~\bibnamefont{Kob}},\ }%
  \bibfield{journal}{%
  \bibinfo {journal} {pressent issue}}%
   (\bibinfo {year} {2014})%
  \bibAnnoteFile{NoStop}{Pedesseau-nbs1}%
\bibitem{vasp_code01}%
  \BibitemOpen
  \bibfield{author}{%
  \bibinfo {author} {\bibfnamefont{G.}~\bibnamefont{Kresse}}\ and\ \bibinfo
  {author} {\bibfnamefont{J.}~\bibnamefont{Furthm\"uller}},\ }%
  \bibfield{journal}{%
  \Doi{10.1103/PhysRevB.54.11169}{\bibinfo {journal} {Phys. Rev. B}}\ }%
  \textbf{\bibinfo {volume} {54}},\ \bibinfo {pages} {11169} (\bibinfo {year}
  {1996})%
  \bibAnnoteFile{NoStop}{vasp_code01}%
\bibitem{vasp_code02}%
  \BibitemOpen
  \bibfield{author}{%
  \bibinfo {author} {\bibfnamefont{G.}~\bibnamefont{Kresse}}\ and\ \bibinfo
  {author} {\bibfnamefont{J.}~\bibnamefont{Furthm\"uller}},\ }%
  \bibfield{journal}{%
  \Doi{10.1016/0927-0256(96)00008-0}{\bibinfo {journal} {Comp. Mat. Science}}\
  }%
  \textbf{\bibinfo {volume} {6}},\ \bibinfo {pages} {15} (\bibinfo {year}
  {1996})%
  \bibAnnoteFile{NoStop}{vasp_code02}%
\bibitem{OMazurin_book}%
  \BibitemOpen
  \bibfield{author}{%
  \bibinfo {author} {\bibfnamefont{O.~V.}\ \bibnamefont{Mazurin}}, \bibinfo
  {author} {\bibfnamefont{T.~P.}\ \bibnamefont{Shvaiko-Shvaikovskaya}},\ and\
  \bibinfo {author} {\bibfnamefont{M.}~\bibnamefont{Streltsina}},\ }%
  \emph{\bibinfo {title} {Handbook of Glass Data: Ternary Silicate Glasses,
  Part C}}\ (\bibinfo {publisher} {Elsevier Science Ltd},\ \bibinfo {year}
  {1984})%
  \bibAnnoteFile{NoStop}{OMazurin_book}%
\bibitem{Kohn1965}%
  \BibitemOpen
  \bibfield{author}{%
  \bibinfo {author} {\bibfnamefont{W.}~\bibnamefont{Kohn}}\ and\ \bibinfo
  {author} {\bibfnamefont{L.~J.}\ \bibnamefont{Sham}},\ }%
  \bibfield{journal}{%
  \Doi{10.1103/PhysRev.140.A1133}{\bibinfo {journal} {Phys. Rev.}}\ }%
  \textbf{\bibinfo {volume} {140}},\ \bibinfo {pages} {A1133} (\bibinfo {year}
  {1965})%
  \bibAnnoteFile{NoStop}{Kohn1965}%
\bibitem{RMartin_book}%
  \BibitemOpen
  \bibfield{author}{%
  \bibinfo {author} {\bibfnamefont{R.}~\bibnamefont{Martin}},\ }%
  \emph{\bibinfo {title} {Electronic Stucture: Basic Theory and Practical
  Methods}}\ (\bibinfo {publisher} {Cambridge University Press},\ \bibinfo
  {year} {2004})%
  \bibAnnoteFile{NoStop}{RMartin_book}%
\bibitem{GGAPBE}%
  \BibitemOpen
  \bibfield{author}{%
  \bibinfo {author} {\bibfnamefont{J.~P.}\ \bibnamefont{Perdew}}, \bibinfo
  {author} {\bibfnamefont{K.}~\bibnamefont{Burke}},\ and\ \bibinfo {author}
  {\bibfnamefont{M.}~\bibnamefont{Ernzerhof}},\ }%
  \bibfield{journal}{%
  \Doi{10.1103/PhysRevLett.77.3865}{\bibinfo {journal} {Phys. Rev. Lett.}}\ }%
  \textbf{\bibinfo {volume} {77}},\ \bibinfo {pages} {3865} (\bibinfo {year}
  {1996})%
  \bibAnnoteFile{NoStop}{GGAPBE}%
\bibitem{PBEsol}%
  \BibitemOpen
  \bibfield{author}{%
  \bibinfo {author} {\bibfnamefont{J.~P.}\ \bibnamefont{Perdew}}, \bibinfo
  {author} {\bibfnamefont{A.}~\bibnamefont{Ruzsinszky}}, \bibinfo {author}
  {\bibfnamefont{G.~I.}\ \bibnamefont{Csonka}}, \bibinfo {author}
  {\bibfnamefont{O.~A.}\ \bibnamefont{Vydrov}}, \bibinfo {author}
  {\bibfnamefont{G.~E.}\ \bibnamefont{Scuseria}}, \bibinfo {author}
  {\bibfnamefont{L.~A.}\ \bibnamefont{Constantin}}, \bibinfo {author}
  {\bibfnamefont{X.}~\bibnamefont{Zhou}},\ and\ \bibinfo {author}
  {\bibfnamefont{K.}~\bibnamefont{Burke}},\ }%
  \bibfield{journal}{%
  \bibinfo {journal} {Phys. Rev. Lett.}\ }%
  \textbf{\bibinfo {volume} {100}},\ \bibinfo {pages} {136406} (\bibinfo {year}
  {2008})%
  \bibAnnoteFile{NoStop}{PBEsol}%
\bibitem{PAW}%
  \BibitemOpen
  \bibfield{author}{%
  \bibinfo {author} {\bibfnamefont{P.~E.}\ \bibnamefont{Bl\"ochl}},\ }%
  \bibfield{journal}{%
  \Doi{10.1103/PhysRevB.50.17953}{\bibinfo {journal} {Phys. Rev. B}}\ }%
  \textbf{\bibinfo {volume} {50}},\ \bibinfo {pages} {17953} (\bibinfo {year}
  {1994})%
  \bibAnnoteFile{NoStop}{PAW}%
\bibitem{Ispas2001}%
  \BibitemOpen
  \bibfield{author}{%
  \bibinfo {author} {\bibfnamefont{S.}~\bibnamefont{Ispas}}, \bibinfo {author}
  {\bibfnamefont{M.}~\bibnamefont{Benoit}}, \bibinfo {author}
  {\bibfnamefont{P.}~\bibnamefont{Jund}},\ and\ \bibinfo {author}
  {\bibfnamefont{R.}~\bibnamefont{Jullien}},\ }%
  \bibfield{journal}{%
  \bibinfo {journal} {Phys. Rev. B}\ }%
  \textbf{\bibinfo {volume} {64}},\ \bibinfo {pages} {214206} (\bibinfo {year}
  {2001})%
  \bibAnnoteFile{NoStop}{Ispas2001}%
\bibitem{Donadio2004214205}%
  \BibitemOpen
  \bibfield{author}{%
  \bibinfo {author} {\bibfnamefont{D.}~\bibnamefont{Donadio}}, \bibinfo
  {author} {\bibfnamefont{M.}~\bibnamefont{Bernasconi}},\ and\ \bibinfo
  {author} {\bibfnamefont{F.}~\bibnamefont{Tassone}},\ }%
  \bibfield{journal}{%
  \bibinfo {journal} {Phys. Rev. B}\ }%
  \textbf{\bibinfo {volume} {70}},\ \bibinfo {pages} {214205} (\bibinfo {year}
  {2004})%
  \bibAnnoteFile{NoStop}{Donadio2004214205}%
\bibitem{Du2006114702}%
  \BibitemOpen
  \bibfield{author}{%
  \bibinfo {author} {\bibfnamefont{J.}~\bibnamefont{Du}}\ and\ \bibinfo
  {author} {\bibfnamefont{L.~R.}\ \bibnamefont{Corrales}},\ }%
  \bibfield{journal}{%
  \bibinfo {journal} {J. Chem. Phys.}\ }%
  \textbf{\bibinfo {volume} {125}},\ \bibinfo {pages} {114702} (\bibinfo {year}
  {2006})%
  \bibAnnoteFile{NoStop}{Du2006114702}%
\bibitem{Ispas2010}%
  \BibitemOpen
  \bibfield{author}{%
  \bibinfo {author} {\bibfnamefont{S.}~\bibnamefont{Ispas}}, \bibinfo {author}
  {\bibfnamefont{T.}~\bibnamefont{Charpentier}}, \bibinfo {author}
  {\bibfnamefont{F.}~\bibnamefont{Mauri}},\ and\ \bibinfo {author}
  {\bibfnamefont{D.~R.}\ \bibnamefont{Neuville}},\ }%
  \bibfield{journal}{%
  \bibinfo {journal} {Sol. St. Sciences}\ }%
  \textbf{\bibinfo {volume} {12}},\ \bibinfo {pages} {183 } (\bibinfo {year}
  {2010})%
  \bibAnnoteFile{NoStop}{Ispas2010}%
\bibitem{Tilocca2008084504}%
  \BibitemOpen
  \bibfield{author}{%
  \bibinfo {author} {\bibfnamefont{A.}~\bibnamefont{Tilocca}},\ }%
  \bibfield{journal}{%
  \bibinfo {journal} {J. Chem. Phys.}\ }%
  \textbf{\bibinfo {volume} {129}},\ \bibinfo {eid} {084504} (\bibinfo {year}
  {2008})%
  \bibAnnoteFile{NoStop}{Tilocca2008084504}%
\bibitem{Vollmayr1996}%
  \BibitemOpen
  \bibfield{author}{%
  \bibinfo {author} {\bibfnamefont{K.}~\bibnamefont{Vollmayr}}, \bibinfo
  {author} {\bibfnamefont{W.}~\bibnamefont{Kob}},\ and\ \bibinfo {author}
  {\bibfnamefont{K.}~\bibnamefont{Binder}},\ }%
  \bibfield{journal}{%
  \bibinfo {journal} {Phys. Rev. B}\ }%
  \textbf{\bibinfo {volume} {54}},\ \bibinfo {pages} {15808} (\bibinfo {year}
  {1996})%
  \bibAnnoteFile{NoStop}{Vollmayr1996}%
\bibitem{Sen199829}%
  \BibitemOpen
  \bibfield{author}{%
  \bibinfo {author} {\bibfnamefont{S.}~\bibnamefont{Sen}}, \bibinfo {author}
  {\bibfnamefont{Z.}~\bibnamefont{Xu}},\ and\ \bibinfo {author}
  {\bibfnamefont{J.~F.}\ \bibnamefont{Stebbins}},\ }%
  \bibfield{journal}{%
  \bibinfo {journal} {J. Non-Cryst. Solids}\ }%
  \textbf{\bibinfo {volume} {226}},\ \bibinfo {pages} {29 } (\bibinfo {year}
  {1998})%
  \bibAnnoteFile{NoStop}{Sen199829}%
\bibitem{Sen199984}%
  \BibitemOpen
  \bibfield{author}{%
  \bibinfo {author} {\bibfnamefont{S.}~\bibnamefont{Sen}},\ }%
  \bibfield{journal}{%
  \bibinfo {journal} {J. Non-Cryst. Solids}\ }%
  \textbf{\bibinfo {volume} {253}},\ \bibinfo {pages} {84 } (\bibinfo {year}
  {1999})%
  \bibAnnoteFile{NoStop}{Sen199984}%
\bibitem{Kiczenski20053571}%
  \BibitemOpen
  \bibfield{author}{%
  \bibinfo {author} {\bibfnamefont{T.}~\bibnamefont{Kiczenski}}, \bibinfo
  {author} {\bibfnamefont{L.-S.}\ \bibnamefont{Du}},\ and\ \bibinfo {author}
  {\bibfnamefont{J.}~\bibnamefont{Stebbins}},\ }%
  \bibfield{journal}{%
  \bibinfo {journal} {J. Non-Cryst. Solids}\ }%
  \textbf{\bibinfo {volume} {351}},\ \bibinfo {pages} {3571 } (\bibinfo {year}
  {2005})%
  \bibAnnoteFile{NoStop}{Kiczenski20053571}%
\bibitem{Angeli2012054110}%
  \BibitemOpen
  \bibfield{author}{%
  \bibinfo {author} {\bibfnamefont{F.}~\bibnamefont{Angeli}}, \bibinfo {author}
  {\bibfnamefont{O.}~\bibnamefont{Villain}}, \bibinfo {author}
  {\bibfnamefont{S.}~\bibnamefont{Schuller}}, \bibinfo {author}
  {\bibfnamefont{T.}~\bibnamefont{Charpentier}}, \bibinfo {author}
  {\bibfnamefont{D.}~\bibnamefont{de~Ligny}}, \bibinfo {author}
  {\bibfnamefont{L.}~\bibnamefont{Bressel}},\ and\ \bibinfo {author}
  {\bibfnamefont{L.}~\bibnamefont{Wondraczek}},\ }%
  \bibfield{journal}{%
  \bibinfo {journal} {Phys. Rev. B}\ }%
  \textbf{\bibinfo {volume} {85}},\ \bibinfo {pages} {054110} (\bibinfo {year}
  {2012})%
  \bibAnnoteFile{NoStop}{Angeli2012054110}%
\bibitem{Majerus2003024210}%
  \BibitemOpen
  \bibfield{author}{%
  \bibinfo {author} {\bibfnamefont{O.}~\bibnamefont{Majerus}}, \bibinfo
  {author} {\bibfnamefont{L.}~\bibnamefont{Cormier}}, \bibinfo {author}
  {\bibfnamefont{G.}~\bibnamefont{Calas}},\ and\ \bibinfo {author}
  {\bibfnamefont{B.}~\bibnamefont{Beuneu}},\ }%
  \bibfield{journal}{%
  \bibinfo {journal} {Phys. Rev. B}\ }%
  \textbf{\bibinfo {volume} {67}},\ \bibinfo {pages} {024210} (\bibinfo {year}
  {2003})%
  \bibAnnoteFile{NoStop}{Majerus2003024210}%
\bibitem{Wu20113944}%
  \BibitemOpen
  \bibfield{author}{%
  \bibinfo {author} {\bibfnamefont{J.}~\bibnamefont{Wu}}, \bibinfo {author}
  {\bibfnamefont{M.}~\bibnamefont{Potuzak}},\ and\ \bibinfo {author}
  {\bibfnamefont{J.~F.}\ \bibnamefont{Stebbins}},\ }%
  \bibfield{journal}{%
  \bibinfo {journal} {J. Non-Cryst. Solids}\ }%
  \textbf{\bibinfo {volume} {357}},\ \bibinfo {pages} {3944 } (\bibinfo {year}
  {2011})%
  \bibAnnoteFile{NoStop}{Wu20113944}%
\bibitem{Sen2007094203}%
  \BibitemOpen
  \bibfield{author}{%
  \bibinfo {author} {\bibfnamefont{S.}~\bibnamefont{Sen}}, \bibinfo {author}
  {\bibfnamefont{T.}~\bibnamefont{Topping}}, \bibinfo {author}
  {\bibfnamefont{P.}~\bibnamefont{Yu}},\ and\ \bibinfo {author}
  {\bibfnamefont{R.~E.}\ \bibnamefont{Youngman}},\ }%
  \bibfield{journal}{%
  \bibinfo {journal} {Phys. Rev. B}\ }%
  \textbf{\bibinfo {volume} {75}},\ \bibinfo {pages} {094203} (\bibinfo {year}
  {2007})%
  \bibAnnoteFile{NoStop}{Sen2007094203}%
\bibitem{Fleet1999233}%
  \BibitemOpen
  \bibfield{author}{%
  \bibinfo {author} {\bibfnamefont{M.}~\bibnamefont{Fleet}}\ and\ \bibinfo
  {author} {\bibfnamefont{S.}~\bibnamefont{Muthupari}},\ }%
  \bibfield{journal}{%
  \bibinfo {journal} {J. Non-Cryst. Solids}\ }%
  \textbf{\bibinfo {volume} {255}},\ \bibinfo {pages} {233 } (\bibinfo {year}
  {1999})%
  \bibAnnoteFile{NoStop}{Fleet1999233}%
\bibitem{Swenson19959310}%
  \BibitemOpen
  \bibfield{author}{%
  \bibinfo {author} {\bibfnamefont{J.}~\bibnamefont{Swenson}}, \bibinfo
  {author} {\bibfnamefont{L.}~\bibnamefont{B\"orjesson}},\ and\ \bibinfo
  {author} {\bibfnamefont{W.~S.}\ \bibnamefont{Howells}},\ }%
  \bibfield{journal}{%
  \bibinfo {journal} {Phys. Rev. B}\ }%
  \textbf{\bibinfo {volume} {52}},\ \bibinfo {pages} {9310} (\bibinfo {year}
  {1995})%
  \bibAnnoteFile{NoStop}{Swenson19959310}%
\bibitem{Kieu20113313}%
  \BibitemOpen
  \bibfield{author}{%
  \bibinfo {author} {\bibfnamefont{L.-H.}\ \bibnamefont{Kieu}}, \bibinfo
  {author} {\bibfnamefont{J.-M.}\ \bibnamefont{Delaye}}, \bibinfo {author}
  {\bibfnamefont{L.}~\bibnamefont{Cormier}},\ and\ \bibinfo {author}
  {\bibfnamefont{C.}~\bibnamefont{Stolz}},\ }%
  \bibfield{journal}{%
  \bibinfo {journal} {J. Non-Cryst. Solids}\ }%
  \textbf{\bibinfo {volume} {357}},\ \bibinfo {pages} {3313 } (\bibinfo {year}
  {2011})%
  \bibAnnoteFile{NoStop}{Kieu20113313}%
\bibitem{Inoue201212325}%
  \BibitemOpen
  \bibfield{author}{%
  \bibinfo {author} {\bibfnamefont{H.}~\bibnamefont{Inoue}}, \bibinfo {author}
  {\bibfnamefont{A.}~\bibnamefont{Masuno}},\ and\ \bibinfo {author}
  {\bibfnamefont{Y.}~\bibnamefont{Watanabe}},\ }%
  \bibfield{journal}{%
  \bibinfo {journal} {J. Phys. Chem. B}\ }%
  \textbf{\bibinfo {volume} {116}},\ \bibinfo {pages} {12325} (\bibinfo {year}
  {2012})%
  \bibAnnoteFile{NoStop}{Inoue201212325}%
\bibitem{Fabian20072084}%
  \BibitemOpen
  \bibfield{author}{%
  \bibinfo {author} {\bibfnamefont{M.}~\bibnamefont{F\'abi\'an}}, \bibinfo
  {author} {\bibfnamefont{E.}~\bibnamefont{Sv\'ab}}, \bibinfo {author}
  {\bibfnamefont{G.}~\bibnamefont{M\'esz\'aros}}, \bibinfo {author}
  {\bibfnamefont{Z.}~\bibnamefont{R\'evay}}, \bibinfo {author}
  {\bibfnamefont{T.}~\bibnamefont{Proffen}},\ and\ \bibinfo {author}
  {\bibfnamefont{E.}~\bibnamefont{Veress}},\ }%
  \bibfield{journal}{%
  \bibinfo {journal} {J. Non-Cryst. Solids}\ }%
  \textbf{\bibinfo {volume} {353}},\ \bibinfo {pages} {2084 } (\bibinfo {year}
  {2007})%
  \bibAnnoteFile{NoStop}{Fabian20072084}%
\bibitem{Wang19991519}%
  \BibitemOpen
  \bibfield{author}{%
  \bibinfo {author} {\bibfnamefont{S.}~\bibnamefont{Wang}}\ and\ \bibinfo
  {author} {\bibfnamefont{J.~F.}\ \bibnamefont{Stebbins}},\ }%
  \bibfield{journal}{%
  \bibinfo {journal} {J. Am. Ceram. Soc.}\ }%
  \textbf{\bibinfo {volume} {82}},\ \bibinfo {pages} {1519} (\bibinfo {year}
  {1999})%
  \bibAnnoteFile{NoStop}{Wang19991519}%
\bibitem{Stebbins199880}%
  \BibitemOpen
  \bibfield{author}{%
  \bibinfo {author} {\bibfnamefont{J.~F.}\ \bibnamefont{Stebbins}}\ and\
  \bibinfo {author} {\bibfnamefont{S.}~\bibnamefont{Sen}},\ }%
  \bibfield{journal}{%
  \bibinfo {journal} {J. Non-Cryst. Solids}\ }%
  \textbf{\bibinfo {volume} {224}},\ \bibinfo {pages} {80 } (\bibinfo {year}
  {1998})%
  \bibAnnoteFile{NoStop}{Stebbins199880}%
\bibitem{Takada19958693}%
  \BibitemOpen
  \bibfield{author}{%
  \bibinfo {author} {\bibfnamefont{A.}~\bibnamefont{Takada}}, \bibinfo {author}
  {\bibfnamefont{C.}~\bibnamefont{Catlow}},\ and\ \bibinfo {author}
  {\bibfnamefont{G.}~\bibnamefont{Price}},\ }%
  \bibfield{journal}{%
  \bibinfo {journal} {Journal of Physics: Condensed Matter}\ }%
  \textbf{\bibinfo {volume} {7}},\ \bibinfo {pages} {8693} (\bibinfo {year}
  {1995})%
  \bibAnnoteFile{NoStop}{Takada19958693}%
\bibitem{Giacomazzi2009064202}%
  \BibitemOpen
  \bibfield{author}{%
  \bibinfo {author} {\bibfnamefont{L.}~\bibnamefont{Giacomazzi}}, \bibinfo
  {author} {\bibfnamefont{P.}~\bibnamefont{Umari}},\ and\ \bibinfo {author}
  {\bibfnamefont{A.}~\bibnamefont{Pasquarello}},\ }%
  \bibfield{journal}{%
  \bibinfo {journal} {Phys. Rev. B}\ }%
  \textbf{\bibinfo {volume} {79}},\ \bibinfo {pages} {064202} (\bibinfo {year}
  {2009})%
  \bibAnnoteFile{NoStop}{Giacomazzi2009064202}%
\bibitem{Benoit2001}%
  \BibitemOpen
  \bibfield{author}{%
  \bibinfo {author} {\bibfnamefont{M.}~\bibnamefont{Benoit}}, \bibinfo {author}
  {\bibfnamefont{S.}~\bibnamefont{Ispas}},\ and\ \bibinfo {author}
  {\bibfnamefont{M.~E.}\ \bibnamefont{Tuckerman}},\ }%
  \bibfield{journal}{%
  \bibinfo {journal} {Phys. Rev. B}\ }%
  \textbf{\bibinfo {volume} {64}},\ \bibinfo {pages} {224205} (\bibinfo {year}
  {2001})%
  \bibAnnoteFile{NoStop}{Benoit2001}%
\bibitem{Tilocca2010014701}%
  \BibitemOpen
  \bibfield{author}{%
  \bibinfo {author} {\bibfnamefont{A.}~\bibnamefont{Tilocca}},\ }%
  \bibfield{journal}{%
  \bibinfo {journal} {J. Chem. Phys.}\ }%
  \textbf{\bibinfo {volume} {133}},\ \bibinfo {eid} {014701} (\bibinfo {year}
  {2010})%
  \bibAnnoteFile{NoStop}{Tilocca2010014701}%
\bibitem{Michel2013169}%
  \BibitemOpen
  \bibfield{author}{%
  \bibinfo {author} {\bibfnamefont{F.}~\bibnamefont{Michel}}, \bibinfo {author}
  {\bibfnamefont{L.}~\bibnamefont{Cormier}}, \bibinfo {author}
  {\bibfnamefont{P.}~\bibnamefont{Lombard}}, \bibinfo {author}
  {\bibfnamefont{B.}~\bibnamefont{Beuneu}}, \bibinfo {author}
  {\bibfnamefont{L.}~\bibnamefont{Galoisy}},\ and\ \bibinfo {author}
  {\bibfnamefont{G.}~\bibnamefont{Calas}},\ }%
  \bibfield{journal}{%
  \bibinfo {journal} {J. Non-Cryst. Solids}\ }%
  \textbf{\bibinfo {volume} {379}},\ \bibinfo {pages} {169 } (\bibinfo {year}
  {2013})%
  \bibAnnoteFile{NoStop}{Michel2013169}%
\bibitem{Ohmura2008224206}%
  \BibitemOpen
  \bibfield{author}{%
  \bibinfo {author} {\bibfnamefont{S.}~\bibnamefont{Ohmura}}\ and\ \bibinfo
  {author} {\bibfnamefont{F.}~\bibnamefont{Shimojo}},\ }%
  \bibfield{journal}{%
  \bibinfo {journal} {Phys. Rev. B}\ }%
  \textbf{\bibinfo {volume} {78}},\ \bibinfo {pages} {224206} (\bibinfo {year}
  {2008})%
  \bibAnnoteFile{NoStop}{Ohmura2008224206}%
\bibitem{Du200310063}%
  \BibitemOpen
  \bibfield{author}{%
  \bibinfo {author} {\bibfnamefont{L.-S.}\ \bibnamefont{Du}}\ and\ \bibinfo
  {author} {\bibfnamefont{J.~F.}\ \bibnamefont{Stebbins}},\ }%
  \bibfield{journal}{%
  \bibinfo {journal} {J. Phys. Chem. B}\ }%
  \textbf{\bibinfo {volume} {107}},\ \bibinfo {pages} {10063} (\bibinfo {year}
  {2003})%
  \bibAnnoteFile{NoStop}{Du200310063}%
\bibitem{Du2004196}%
  \BibitemOpen
  \bibfield{author}{%
  \bibinfo {author} {\bibfnamefont{L.-S.}\ \bibnamefont{Du}}, \bibinfo {author}
  {\bibfnamefont{J.}~\bibnamefont{Allwardt}}, \bibinfo {author}
  {\bibfnamefont{B.}~\bibnamefont{Schmidt}},\ and\ \bibinfo {author}
  {\bibfnamefont{J.}~\bibnamefont{Stebbins}},\ }%
  \bibfield{journal}{%
  \bibinfo {journal} {J. Non-Cryst. Solids}\ }%
  \textbf{\bibinfo {volume} {337}},\ \bibinfo {pages} {196 } (\bibinfo {year}
  {2004})%
  \bibAnnoteFile{NoStop}{Du2004196}%
\bibitem{Umari2005137401}%
  \BibitemOpen
  \bibfield{author}{%
  \bibinfo {author} {\bibfnamefont{P.}~\bibnamefont{Umari}}\ and\ \bibinfo
  {author} {\bibfnamefont{A.}~\bibnamefont{Pasquarello}},\ }%
  \bibfield{journal}{%
  \bibinfo {journal} {Phys. Rev. Lett.}\ }%
  \textbf{\bibinfo {volume} {95}},\ \bibinfo {pages} {137401} (\bibinfo {year}
  {2005})%
  \bibAnnoteFile{NoStop}{Umari2005137401}%
\bibitem{Ferlat2008065504}%
  \BibitemOpen
  \bibfield{author}{%
  \bibinfo {author} {\bibfnamefont{G.}~\bibnamefont{Ferlat}}, \bibinfo {author}
  {\bibfnamefont{T.}~\bibnamefont{Charpentier}}, \bibinfo {author}
  {\bibfnamefont{A.~P.}\ \bibnamefont{Seitsonen}}, \bibinfo {author}
  {\bibfnamefont{A.}~\bibnamefont{Takada}}, \bibinfo {author}
  {\bibfnamefont{M.}~\bibnamefont{Lazzeri}}, \bibinfo {author}
  {\bibfnamefont{L.}~\bibnamefont{Cormier}}, \bibinfo {author}
  {\bibfnamefont{G.}~\bibnamefont{Calas}},\ and\ \bibinfo {author}
  {\bibfnamefont{F.}~\bibnamefont{Mauri}},\ }%
  \bibfield{journal}{%
  \bibinfo {journal} {Phys. Rev. Lett.}\ }%
  \textbf{\bibinfo {volume} {101}},\ \bibinfo {pages} {065504} (\bibinfo {year}
  {2008})%
  \bibAnnoteFile{NoStop}{Ferlat2008065504}%
\bibitem{Ispas2005}%
  \BibitemOpen
  \bibfield{author}{%
  \bibinfo {author} {\bibfnamefont{S.}~\bibnamefont{Ispas}}, \bibinfo {author}
  {\bibfnamefont{N.}~\bibnamefont{Zotov}}, \bibinfo {author}
  {\bibfnamefont{S.}~\bibnamefont{De~Wispelaere}},\ and\ \bibinfo {author}
  {\bibfnamefont{W.}~\bibnamefont{Kob}},\ }%
  \bibfield{journal}{%
  \bibinfo {journal} {J. Non-Cryst. Solids}\ }%
  \textbf{\bibinfo {volume} {351}},\ \bibinfo {pages} {1144 } (\bibinfo {year}
  {2005})%
  \bibAnnoteFile{NoStop}{Ispas2005}%
\bibitem{Taraskin1997}%
  \BibitemOpen
  \bibfield{author}{%
  \bibinfo {author} {\bibfnamefont{S.~N.}\ \bibnamefont{Taraskin}}\ and\
  \bibinfo {author} {\bibfnamefont{S.~R.}\ \bibnamefont{Elliott}},\ }%
  \bibfield{journal}{%
  \bibinfo {journal} {Phys. Rev. B}\ }%
  \textbf{\bibinfo {volume} {56}},\ \bibinfo {pages} {8605} (\bibinfo {year}
  {1997})%
  \bibAnnoteFile{NoStop}{Taraskin1997}%
\bibitem{Pasquarello1998}%
  \BibitemOpen
  \bibfield{author}{%
  \bibinfo {author} {\bibfnamefont{A.}~\bibnamefont{Pasquarello}}, \bibinfo
  {author} {\bibfnamefont{J.}~\bibnamefont{Sarnthein}},\ and\ \bibinfo {author}
  {\bibfnamefont{R.}~\bibnamefont{Car}},\ }%
  \bibfield{journal}{%
  \bibinfo {journal} {Phys. Rev. B}\ }%
  \textbf{\bibinfo {volume} {57}},\ \bibinfo {pages} {14133} (\bibinfo {year}
  {1998})%
  \bibAnnoteFile{NoStop}{Pasquarello1998}%
\bibitem{DeSousaMeneses201350}%
  \BibitemOpen
  \bibfield{author}{%
  \bibinfo {author} {\bibfnamefont{D.~D.~S.}\ \bibnamefont{Meneses}}, \bibinfo
  {author} {\bibfnamefont{M.}~\bibnamefont{Eckes}}, \bibinfo {author}
  {\bibfnamefont{L.}~\bibnamefont{del Campo}}, \bibinfo {author}
  {\bibfnamefont{C.~N.}\ \bibnamefont{Santos}}, \bibinfo {author}
  {\bibfnamefont{Y.}~\bibnamefont{Vaills}},\ and\ \bibinfo {author}
  {\bibfnamefont{P.}~\bibnamefont{Echegut}},\ }%
  \bibfield{journal}{%
  \bibinfo {journal} {Vibrational Spectroscopy}\ }%
  \textbf{\bibinfo {volume} {65}},\ \bibinfo {pages} {50 } (\bibinfo {year}
  {2013})%
  \bibAnnoteFile{NoStop}{DeSousaMeneses201350}%
\bibitem{Yano2003137}%
  \BibitemOpen
  \bibfield{author}{%
  \bibinfo {author} {\bibfnamefont{T.}~\bibnamefont{Yano}}, \bibinfo {author}
  {\bibfnamefont{N.}~\bibnamefont{Kunimine}}, \bibinfo {author}
  {\bibfnamefont{S.}~\bibnamefont{Shibata}},\ and\ \bibinfo {author}
  {\bibfnamefont{M.}~\bibnamefont{Yamane}},\ }%
  \bibfield{journal}{%
  \bibinfo {journal} {J. Non-Cryst. Solids}\ }%
  \textbf{\bibinfo {volume} {321}},\ \bibinfo {pages} {137 } (\bibinfo {year}
  {2003})%
  \bibAnnoteFile{NoStop}{Yano2003137}%
\bibitem{Simon2008155103}%
  \BibitemOpen
  \bibfield{author}{%
  \bibinfo {author} {\bibfnamefont{G.}~\bibnamefont{Simon}}, \bibinfo {author}
  {\bibfnamefont{B.}~\bibnamefont{Hehlen}}, \bibinfo {author}
  {\bibfnamefont{R.}~\bibnamefont{Vacher}},\ and\ \bibinfo {author}
  {\bibfnamefont{E.}~\bibnamefont{Courtens}},\ }%
  \bibfield{journal}{%
  \bibinfo {journal} {J. Phys.: Condensed Matter}\ }%
  \textbf{\bibinfo {volume} {20}},\ \bibinfo {pages} {155103} (\bibinfo {year}
  {2008})%
  \bibAnnoteFile{NoStop}{Simon2008155103}%
\bibitem{Varsamis19993885}%
  \BibitemOpen
  \bibfield{author}{%
  \bibinfo {author} {\bibfnamefont{C.~P.}\ \bibnamefont{Varsamis}}, \bibinfo
  {author} {\bibfnamefont{E.~I.}\ \bibnamefont{Kamitsos}},\ and\ \bibinfo
  {author} {\bibfnamefont{G.~D.}\ \bibnamefont{Chryssikos}},\ }%
  \bibfield{journal}{%
  \bibinfo {journal} {Phys. Rev. B}\ }%
  \textbf{\bibinfo {volume} {60}},\ \bibinfo {pages} {3885} (\bibinfo {year}
  {1999})%
  \bibAnnoteFile{NoStop}{Varsamis19993885}%
\bibitem{Sarnthein199512690}%
  \BibitemOpen
  \bibfield{author}{%
  \bibinfo {author} {\bibfnamefont{J.}~\bibnamefont{Sarnthein}}, \bibinfo
  {author} {\bibfnamefont{A.}~\bibnamefont{Pasquarello}},\ and\ \bibinfo
  {author} {\bibfnamefont{R.}~\bibnamefont{Car}},\ }%
  \bibfield{journal}{%
  \bibinfo {journal} {Phys. Rev. B}\ }%
  \textbf{\bibinfo {volume} {52}},\ \bibinfo {pages} {12690} (\bibinfo {year}
  {1995})%
  \bibAnnoteFile{NoStop}{Sarnthein199512690}%
\bibitem{Benoit2000}%
  \BibitemOpen
  \bibfield{author}{%
  \bibinfo {author} {\bibnamefont{{M. Benoit}}}, \bibinfo {author}
  {\bibnamefont{{S. Ispas}}}, \bibinfo {author} {\bibnamefont{{P. Jund}}},\
  and\ \bibinfo {author} {\bibnamefont{{R. Jullien}}},\ }%
  \bibfield{journal}{%
  \bibinfo {journal} {Eur. Phys. J. B}\ }%
  \textbf{\bibinfo {volume} {13}},\ \bibinfo {pages} {631} (\bibinfo {year}
  {2000})%
  \bibAnnoteFile{NoStop}{Benoit2000}%
\bibitem{Burke2012150901}%
  \BibitemOpen
  \bibfield{author}{%
  \bibinfo {author} {\bibfnamefont{K.}~\bibnamefont{Burke}},\ }%
  \bibfield{journal}{%
  \bibinfo {journal} {J. Chem. Phys.}\ }%
  \textbf{\bibinfo {volume} {136}},\ \bibinfo {pages} {150901} (\bibinfo {year}
  {2012})%
  \bibAnnoteFile{NoStop}{Burke2012150901}%
\bibitem{Du200622346}%
  \BibitemOpen
  \bibfield{author}{%
  \bibinfo {author} {\bibfnamefont{J.}~\bibnamefont{Du}}\ and\ \bibinfo
  {author} {\bibfnamefont{L.~R.}\ \bibnamefont{Corrales}},\ }%
  \bibfield{journal}{%
  \bibinfo {journal} {J. Phys. Chem.}\ }%
  \textbf{\bibinfo {volume} {110}},\ \bibinfo {pages} {22346} (\bibinfo {year}
  {2006})%
  \bibAnnoteFile{NoStop}{Du200622346}%
\bibitem{Bader_book}%
  \BibitemOpen
  \bibfield{author}{%
  \bibinfo {author} {\bibfnamefont{R.~F.~W.}\ \bibnamefont{Bader}},\ }%
  \emph{\bibinfo {title} {Atoms in Molecule: A Quantum Theory}}\ (\bibinfo
  {publisher} {Oxford University Press, Oxford},\ \bibinfo {year} {1990})%
  \bibAnnoteFile{NoStop}{Bader_book}%
\bibitem{Henkelman2006}%
  \BibitemOpen
  \bibfield{author}{%
  \bibinfo {author} {\bibfnamefont{G.}~\bibnamefont{Henkelman}}, \bibinfo
  {author} {\bibfnamefont{A.}~\bibnamefont{Arnaldsson}},\ and\ \bibinfo
  {author} {\bibfnamefont{J.}~\bibnamefont{H.}},\ }%
  \bibfield{journal}{%
  \bibinfo {journal} {Comput. Mater. Sci.}\ }%
  \textbf{\bibinfo {volume} {36}},\ \bibinfo {pages} {354 } (\bibinfo {year}
  {2006})%
  \bibAnnoteFile{NoStop}{Henkelman2006}%
\bibitem{Bruckner198049}%
  \BibitemOpen
  \bibfield{author}{%
  \bibinfo {author} {\bibfnamefont{R.}~\bibnamefont{Bruckner}}, \bibinfo
  {author} {\bibfnamefont{H.-U.}\ \bibnamefont{Chun}}, \bibinfo {author}
  {\bibfnamefont{H.}~\bibnamefont{Goretzki}},\ and\ \bibinfo {author}
  {\bibfnamefont{M.}~\bibnamefont{Sammet}},\ }%
  \bibfield{journal}{%
  \bibinfo {journal} {J. Non-Cryst. Solids}\ }%
  \textbf{\bibinfo {volume} {42}},\ \bibinfo {pages} {49 } (\bibinfo {year}
  {1980})%
  \bibAnnoteFile{NoStop}{Bruckner198049}%
\bibitem{Uchino1991}%
  \BibitemOpen
  \bibfield{author}{%
  \bibinfo {author} {\bibfnamefont{T.}~\bibnamefont{Uchino}}, \bibinfo {author}
  {\bibfnamefont{M.}~\bibnamefont{Iwasaki}}, \bibinfo {author}
  {\bibfnamefont{T.}~\bibnamefont{Sakka}},\ and\ \bibinfo {author}
  {\bibfnamefont{Y.}~\bibnamefont{Ogata}},\ }%
  \bibfield{journal}{%
  \bibinfo {journal} {J. Phys. Chem.}\ }%
  \textbf{\bibinfo {volume} {95}},\ \bibinfo {pages} {5455} (\bibinfo {year}
  {1991})%
  \bibAnnoteFile{NoStop}{Uchino1991}%
\bibitem{Zwijnenburg2002}%
  \BibitemOpen
  \bibfield{author}{%
  \bibinfo {author} {\bibfnamefont{M.~A.}\ \bibnamefont{Zwijnenburg}}, ,
  \bibinfo {author} {\bibfnamefont{C.}~\bibnamefont{van Alsenoy}},\ and\
  \bibinfo {author} {\bibfnamefont{T.}~\bibnamefont{Maschmeyer}},\ }%
  \bibfield{journal}{%
  \bibinfo {journal} {J. Phys. Chem. A}\ }%
  \textbf{\bibinfo {volume} {106}},\ \bibinfo {pages} {12376} (\bibinfo {year}
  {2002})%
  \bibAnnoteFile{NoStop}{Zwijnenburg2002}%
\bibitem{Gonze199710335}%
  \BibitemOpen
  \bibfield{author}{%
  \bibinfo {author} {\bibfnamefont{X.}~\bibnamefont{Gonze}}\ and\ \bibinfo
  {author} {\bibfnamefont{C.}~\bibnamefont{Lee}},\ }%
  \bibfield{journal}{%
  \bibinfo {journal} {Phys. Rev. B}\ }%
  \textbf{\bibinfo {volume} {55}},\ \bibinfo {pages} {10355} (\bibinfo {year}
  {1997})%
  \bibAnnoteFile{NoStop}{Gonze199710335}%
\bibitem{Pasquarello19971766}%
  \BibitemOpen
  \bibfield{author}{%
  \bibinfo {author} {\bibfnamefont{A.}~\bibnamefont{Pasquarello}}\ and\
  \bibinfo {author} {\bibfnamefont{R.}~\bibnamefont{Car}},\ }%
  \bibfield{journal}{%
  \bibinfo {journal} {Phys. Rev. Lett.}\ }%
  \textbf{\bibinfo {volume} {79}},\ \bibinfo {pages} {1766} (\bibinfo {year}
  {1997})%
  \bibAnnoteFile{NoStop}{Pasquarello19971766}%
\bibitem{Gajdos2006045112}%
  \BibitemOpen
  \bibfield{author}{%
  \bibinfo {author} {\bibfnamefont{M.}~\bibnamefont{Gajdo\v{s}}}, \bibinfo
  {author} {\bibfnamefont{K.}~\bibnamefont{Hummer}}, \bibinfo {author}
  {\bibfnamefont{G.}~\bibnamefont{Kresse}}, \bibinfo {author}
  {\bibfnamefont{J.}~\bibnamefont{Furthm\"uller}},\ and\ \bibinfo {author}
  {\bibfnamefont{F.}~\bibnamefont{Bechstedt}},\ }%
  \bibfield{journal}{%
  \bibinfo {journal} {Phys. Rev. B}\ }%
  \textbf{\bibinfo {volume} {73}},\ \bibinfo {pages} {045112} (\bibinfo {year}
  {2006})%
  \bibAnnoteFile{NoStop}{Gajdos2006045112}%
\bibitem{Philipp1998}%
  \BibitemOpen
  \bibfield{author}{%
  \bibinfo {author} {\bibfnamefont{H.}~\bibnamefont{Philipp}},\ }%
  \enquote{\bibinfo {title} {Silicon dioxide (sio$_2$) (glass)},}\ in\
  \emph{\bibinfo {booktitle} {Handbook of Optical Constants of Solids}}\
  (\bibinfo {publisher} {ed. D. Palik, Academic Press, San Diego},\ \bibinfo
  {year} {1998})\ pp.\ \bibinfo {pages} {749--763}%
  \bibAnnoteFile{NoStop}{Philipp1998}%
\bibitem{Kamitsos199431}%
  \BibitemOpen
  \bibfield{author}{%
  \bibinfo {author} {\bibfnamefont{E.~I.}\ \bibnamefont{Kamitsos}}, \bibinfo
  {author} {\bibfnamefont{J.~A.}\ \bibnamefont{Kapoutsis}}, \bibinfo {author}
  {\bibfnamefont{H.}~\bibnamefont{Jain}},\ and\ \bibinfo {author}
  {\bibfnamefont{C.~H.}\ \bibnamefont{Hsieh}},\ }%
  \bibfield{journal}{%
  \bibinfo {journal} {J. Non-Cryst. Solids}\ }%
  \textbf{\bibinfo {volume} {171}},\ \bibinfo {pages} {31} (\bibinfo {year}
  {1994})%
  \bibAnnoteFile{NoStop}{Kamitsos199431}%
\bibitem{Hsieh19961704}%
  \BibitemOpen
  \bibfield{author}{%
  \bibinfo {author} {\bibfnamefont{C.~H.}\ \bibnamefont{Hsieh}}, \bibinfo
  {author} {\bibfnamefont{H.}~\bibnamefont{Jain}},\ and\ \bibinfo {author}
  {\bibfnamefont{E.~I.}\ \bibnamefont{Kamitsos}},\ }%
  \bibfield{journal}{%
  \bibinfo {journal} {J. Appl. Phys.}\ }%
  \textbf{\bibinfo {volume} {80}},\ \bibinfo {pages} {1704 } (\bibinfo {year}
  {1996})%
  \bibAnnoteFile{NoStop}{Hsieh19961704}%
\bibitem{Wang20081128}%
  \BibitemOpen
  \bibfield{author}{%
  \bibinfo {author} {\bibfnamefont{Z.}~\bibnamefont{Wang}}, \bibinfo {author}
  {\bibfnamefont{Y.}~\bibnamefont{Hu}}, \bibinfo {author}
  {\bibfnamefont{H.}~\bibnamefont{Lu}},\ and\ \bibinfo {author}
  {\bibfnamefont{F.}~\bibnamefont{Yu}},\ }%
  \bibfield{journal}{%
  \bibinfo {journal} {J. Non-Cryst. Solids}\ }%
  \textbf{\bibinfo {volume} {354}},\ \bibinfo {pages} {1128 } (\bibinfo {year}
  {2008})%
  \bibAnnoteFile{NoStop}{Wang20081128}%
\bibitem{Thorpe19868490}%
  \BibitemOpen
  \bibfield{author}{%
  \bibinfo {author} {\bibfnamefont{M.~F.}\ \bibnamefont{Thorpe}}\ and\ \bibinfo
  {author} {\bibfnamefont{S.~W.}\ \bibnamefont{de~Leeuw}},\ }%
  \bibfield{journal}{%
  \bibinfo {journal} {Phys. Rev. B}\ }%
  \textbf{\bibinfo {volume} {33}},\ \bibinfo {pages} {8490} (\bibinfo {year}
  {1986})%
  \bibAnnoteFile{NoStop}{Thorpe19868490}%
\bibitem{GaleenerFL}%
  \BibitemOpen
  \bibfield{author}{%
  \bibinfo {author} {\bibfnamefont{F.~L.}\ \bibnamefont{Galeener}}, \bibinfo
  {author} {\bibfnamefont{G.}~\bibnamefont{Lucovsky}},\ and\ \bibinfo {author}
  {\bibfnamefont{J.~C.}\ \bibnamefont{Mikkelsen}},\ }%
  \bibfield{journal}{%
  \Doi{10.1103/PhysRevB.22.3983}{\bibinfo {journal} {Phys. Rev. B}}\ }%
  \textbf{\bibinfo {volume} {22}},\ \bibinfo {pages} {3983} (\bibinfo {year}
  {1980})%
  \bibAnnoteFile{NoStop}{GaleenerFL}%
\bibitem{FDomine01}%
  \BibitemOpen
  \bibfield{author}{%
  \bibinfo {author} {\bibfnamefont{F.}~\bibnamefont{Domine}}\ and\ \bibinfo
  {author} {\bibfnamefont{B.}~\bibnamefont{Piriou}},\ }%
  \bibfield{journal}{%
  \Doi{10.1016/0022-3093(83)90012-1}{\bibinfo {journal} {J. Non-Cryst.
  Solids}}\ }%
  \textbf{\bibinfo {volume} {55}},\ \bibinfo {pages} {125} (\bibinfo {year}
  {1983})%
  \bibAnnoteFile{NoStop}{FDomine01}%
\bibitem{CMerzbacher01}%
  \BibitemOpen
  \bibfield{author}{%
  \bibinfo {author} {\bibfnamefont{C.~I.}\ \bibnamefont{Merzbacher}}\ and\
  \bibinfo {author} {\bibfnamefont{W.~B.}\ \bibnamefont{White}},\ }%
  \bibfield{journal}{%
  \bibinfo {journal} {American Mineralogist}\ }%
  \textbf{\bibinfo {volume} {73}},\ \bibinfo {pages} {1089} (\bibinfo {year}
  {1988})%
  \bibAnnoteFile{NoStop}{CMerzbacher01}%
\bibitem{Kamitsos199759}%
  \BibitemOpen
  \bibfield{author}{%
  \bibinfo {author} {\bibfnamefont{E.}~\bibnamefont{Kamitsos}}, \bibinfo
  {author} {\bibfnamefont{Y.}~\bibnamefont{Yiannopoulos}}, \bibinfo {author}
  {\bibfnamefont{C.}~\bibnamefont{Varsamis}},\ and\ \bibinfo {author}
  {\bibfnamefont{H.}~\bibnamefont{Jain}},\ }%
  \bibfield{journal}{%
  \bibinfo {journal} {J. Non-Cryst. Solids}\ }%
  \textbf{\bibinfo {volume} {222}},\ \bibinfo {pages} {59 } (\bibinfo {year}
  {1997})%
  \bibAnnoteFile{NoStop}{Kamitsos199759}%
\end{thebibliography}
\end{document}